\documentclass[12pt]{article}
\usepackage{epsfig}
\newcommand{\nn}{\nonumber}
\newcommand{\raw}{\rightarrow}

\newcommand{\be}{\begin{equation}}
\newcommand{\ee}{\end{equation}}
\newcommand{\bea}{\begin{eqnarray}}
\newcommand{\eea}{\end{eqnarray}}

\def\beq{\begin{equation}}
\def\eeq{\end{equation}}

\newcommand{\tetaot}{\mbox{$\theta_{13}$}}
\newcommand{\tetatt}{\mbox{$\theta_{23}$}}
\newcommand{\deltt}{\mbox{$\Delta_{23}$}}
\newcommand{\delot}{\mbox{$\Delta_{13}$}}

\newcommand{\numu}{\mbox{$\nu_{\mu}$}}
\newcommand{\numubar}{\mbox{$\overline{\nu}_{\mu}$}}
\newcommand{\nue}{\mbox{$\nu_{e}$}}
\newcommand{\nuebar}{\mbox{$\overline{\nu}_{e}$}}

\begin{document}
%
%
\thispagestyle{empty}
\begin{flushright}
{\tt hep-ph/0002108}\\
{CERN-TH/2000-40}\\
{FTUAM-00-03}\\
{IFT-UAM/CSIC-00-04} \\
{FTUV/00-12}\\
{IFIC/00-13}
\end{flushright}
\vspace*{1cm}
\begin{center}
{\Large{\bf Golden measurements at a neutrino factory} }\\
\vspace{.5cm}
A. Cervera$^{\rm a,}$\footnote{anselmo.cervera@cern.ch},
A. Donini$^{\rm b,}$\footnote{donini@daniel.ft.uam.es},
M.B. Gavela$^{\rm b,}$\footnote{gavela@garuda.ft.uam.es},
J.J. Gomez C\'adenas$^{\rm a,}$\footnote{gomez@hal.ific.uv.es},
P. Hern\'andez$^{\rm c,}$\footnote{pilar.hernandez@cern.ch. On leave 
from Dept. de F\'{\i}sica Te\'orica, Universidad de Valencia.},
O. Mena$^{\rm b,}$\footnote{mena@delta.ft.uam.es}
and S. Rigolin$^{\rm d,}$\footnote{srigolin@umich.edu}
 
\vspace*{1cm}
$^{\rm a}$ Dept. de F\'{\i}sica At\'omica y Nuclear and IFIC, Universidad de Valencia, 
Spain \\
$^{\rm b}$ Dept. de F\'{\i}sica Te\'orica, Univ. Aut\'onoma de
Madrid, 28049 Spain \\
$^{\rm c}$ Theory Division, CERN, 1211 Geneva 23, Switzerland \\
$^{\rm d}$ Dept. of Physics, University of Michigan, Ann Arbor, MI 48105 USA

\end{center}
\vspace{.3cm}
\begin{abstract}
\noindent

The precision and discovery potential of a neutrino factory based 
on muon storage rings is studied. 
For three-family neutrino oscillations, we analyse how to measure or 
severely constraint the angle $\theta_{13}$, CP violation, MSW effects 
and the sign of the atmospheric mass difference $\Delta m^2_{23}$. 
We present a simple analytical formula for the oscillation probabilities 
in matter, with all neutrino mass differences non-vanishing, which clarifies 
the subtleties involved in disentangling the unknown parameters. 
The appearance of ``wrong-sign muons'' at three reference baselines 
is considered: 732 km, 3500 km, and 7332 km. We exploit the dependence of
the signal on the neutrino energy, and include as well realistic background 
estimations and detection efficiencies. 
The optimal baseline turns out to be ${\cal O}(3000$ km).
Analyses combining the information from different baselines are also 
presented.  
\end{abstract}

\newpage
\pagestyle{plain} 
\setcounter{page}{1}
\setcounter{footnote}{0}

%
%
\section{Introduction}
%

The atmospheric \cite{Superka,otheratm} plus solar \cite{sol} 
neutrino data point to neutrino oscillations \cite{osc,MSW} and can be 
easily accommodated in a three-family mixing scenario.

Let $ U$, with $(\nu_e,\nu_\mu,\nu_\tau)^T= U \cdot (\nu_1,\nu_2,\nu_3)^T$, 
be the leptonic Cabibbo-Kobayashi-Maskawa (CKM) matrix in its most conventional
parametrization~\cite{PDG}:
\begin{equation}
 U \equiv  U_{23} U_{13} U_{12} \equiv
\left(\matrix{         1 &        0 &       0              \cr 
                       0 &   c_{23} &  s_{23}              \cr
                       0 & - s_{23} &  c_{23}              \cr } \right) \;
\left(\matrix{    c_{13} &        0 &  s_{13} e^{i \delta} \cr
                       0 &        1 &       0              \cr
 -  s_{13} e^{-i \delta} &        0 &  c_{13}              \cr } \right) \;
\left(\matrix{    c_{12} &   s_{12} &       0              \cr
                - s_{12} &   c_{12} &       0              \cr
                       0 &        0 &       1              \cr } \right) 
\label{CKM}
\end{equation}
with $s_{12} \equiv \sin \theta_{12}$, and similarly for the other 
sines and cosines.  
Oscillation experiments are sensitive to the neutrino mass differences and the 
four parameters in the mixing matrix of Eq.~(\ref{CKM}): three angles 
and the Dirac CP-odd phase. 

The SuperKamiokande \cite{Superka} data on atmospheric neutrinos 
are interpreted as oscillations of muon neutrinos into
neutrinos that are not $\nu_e$'s, with a mass gap that we denote\footnote{ 
$\Delta m^2_{ij} \equiv m^2_j - m^2_i$ throughout the paper.} 
by $\Delta m^2_{23}$. Roughly speaking, the measured mixing angle $\theta_{23}$ is 
close to maximal and $|\Delta m^2_{23}|$ 
is in the range $10^{-3}$--$10^{-2}$ eV$^2$. 
The solar neutrino deficit is interpreted either as MSW (matter
enhanced) oscillations \cite{MSW} or as vacuum oscillations (VO) \cite{osc} 
that deplete the original $\nu_e$'s, presumably in favour of $\nu_\mu$'s 
or alternatively into sterile neutrinos.
The corresponding squared mass differences --${\cal O} (10^{-5}$-$10^{-4})$ eV$^2$ 
for the large mixing angle MSW solution (LMA-MSW), ${\cal O} (10^{-6})$ eV$^2$ for the 
small mixing angle MSW solution (SMA-MSW), or ${\cal O} (10^{-10})$ eV$^2$ for VO-- 
are significantly below the range deduced from atmospheric observations. 
We identify this mass difference with $\Delta m^2_{12}$ in this 
parametrization.  Its sign is constrained by solar data: while the SMA-MSW 
solution exists only for positive
 $\Delta m^2_{12}$, in the LMA-MSW range there is also a small window
 at negative values \cite{dfm}. 

These oscillation signals will be confirmed and further constrained in
ongoing and planned atmospheric, solar and long baseline reactor 
experiments \cite{futuresolar}, as well as in future long baseline 
accelerator neutrino experiments \cite{futurelbl}. 
In a few years they will answer the question of sterile neutrinos contributing or not 
to present data. The MSW effect is expected to play a major role in explaining the solar deficit 
and both solar and reactor experiments will also clarify whether Nature has chosen 
the LMA-MSW rather than SMA-MSW or VO solutions.

The atmospheric neutrino parameters will be known  with better precision as 
well.
Experimental information relevant for a more precise knowledge 
of the atmospheric neutrino fluxes will be available \cite{dydak,ting}.  
Also, projected long baseline accelerator experiments will improve the precision of $|\Delta m^2_{23}|$ and $\theta_{23}$. For instance, $|\Delta m^2_{23}|$ is expected 
to be measured at MINOS with an accuracy below $10\%$ if 
$|\Delta m^2_{23}|> 3\times 10^{-3}$ eV$^2$ \cite{petyt}. 

Nevertheless, there is a strong case for going further in the fundamental quest 
of the neutrino masses and mixing angles, as a necessary step to unravel the 
fundamental new scale(s) behind neutrino oscillations.
In ten years from now no significant improvement is expected  
in the knowledge of:

\begin{itemize}

\item 
The angle $\tetaot$, which is the key between the atmospheric and solar neutrino realms, 
for which the present CHOOZ bound is $\sin^2 \tetaot \le 5\times 10^{-2}$ \cite{chooz}\cite{concha}. 

\item 
The sign of $\Delta m^2_{23}$, which 
determines whether the three-family neutrino spectrum is of the ``hierarchical''
or ``degenerate''  type (i.e. only one heavy state and two almost degenerate 
light ones, or the reverse).

\item 
Leptonic CP-violation.

\item 
The precise study of matter effects in the $\nu$ propagation through the Earth:
a model-independent experimental confirmation of the MSW effect will not be available.

\end{itemize}

The most sensitive method to study these topics is to measure the transition probabilities 
involving $\nu_e$ and $\bar \nu_e$, in particular 
$\nu_e(\bar \nu_e) \raw \nu_\mu ( \bar \nu_\mu)$. 
This is precisely the golden measurement at the {\it neutrino factory} \cite{geer}. 
Such a facility is unique in providing high energy and intense $\nu_e (\bar \nu_e)$ 
beams coming from positive (negative) muons which decay in the straight sections 
of a muon storage ring \cite{firstmachine}. 
Since these beams contain also $\bar \nu_\mu (\nu_\mu)$ (but no $\nu_\mu (\bar \nu_\mu)$!), the transitions
of interest can be measured by searching for ``wrong-sign'' muons: 
negative (positive) muons  appearing in a massive detector with good muon charge 
identification capabilities.

The first exploratory studies of the use of a neutrino beam with these
characteristics \cite{usaworkshop} were done in the context of 
two-family mixing. In this approximation, the wrong-sign muon signal 
in the atmospheric range is absent, since the atmospheric oscillation is
$\nu_\mu \leftrightarrow \nu_\tau$. The enormous physics reach of such signals 
in the context of three-family neutrino mixing was first realized in \cite{dgh}, 
where the authors put the emphasis on the measurement of the angle $\tetaot$ and 
CP-violation (see also \cite{bruno}). The latter may be at reach if the solar deficit 
corresponds to the LMA-MSW solution \cite{dgh, romanino, donini}.
Recently, it has also been shown \cite{bgrw} that the precision in the 
knowledge of the atmospheric parameters $\tetatt$ and $|\Delta m^2_{23}|$ 
can reach the percent level at a neutrino factory, using muon disappearance measurements. 
Furthermore, it was pointed out \cite{Lipari:2000wy,ellis,bgrw} that the sign 
of $\Delta m^2_{23}$ 
can also be determined at long baselines, through sizeable matter 
effects. The importance of measuring precisely Earth matter 
effects in a clean and model-independent way, both for the understanding of 
the fundamental parameters and for their intrinsic interest has been 
stressed in refs. \cite{bcr,bgw,Mocioiu:1999ag,Petcov}. 

The aim of this paper is to identify the optimal baselines for studying 
the above itemized topics. This requires to include in the analysis the maximum information 
that can be attained at any fixed baseline. While all previous analyses 
have been based in energy-integrated quantities, we will take into account the 
neutrino energy dependence of the wrong-sign muon signals, 
together with the information obtained from running in the two different 
beam polarities. 

We will consider in turn scenarios in which the solar oscillation lies in 
the SMA-MSW or VO range 
and in the LMA-MSW range. In the latter, the dependence of the oscillation probabilities on the 
solar parameters $\theta_{12}, \Delta m^2_{12}$, and on the CP-odd phase, 
$\delta$,  
is sizeable at 
terrestrial distances and complicates the 
measurement of $\tetaot$ due to the 
presence of other 
unknowns (mainly $\delta$). This potential 
difficulty was first pointed out in \cite{romanino}. The previous analysis of 
the sensitivity to $\tetaot$ \cite{dgh} neglected solar parameters 
and is thus only valid for the SMA-MSW and VO solutions, or  
 the LMA-MSW if $\tetaot$ is large enough.
In the present paper a higher statistics is considered, allowing to explore
 smaller values of $\tetaot$, and the remark is very pertinent. 
We will discuss in detail the 
issue of how to disentangle  $\tetaot$ and 
$\delta$, guided by an approximate analytical formula for the oscillation probabilities in matter
including two distinct mass differences. As we will see, the choice of the correct baseline 
is essential to solve this problem. 

We shall consider the following ``reference set-up'': neutrino beams resulting from 
the decay of $ 2 \times 10^{20} \mu^+$'s and/or $\mu^-$'s per year in a straight section 
of an $E_\mu = 50$ GeV muon accumulator. A long baseline (LBL) experiment with a 40 kT 
detector and five years of data taking for each polarity is considered. 
Alternatively, the same results could be obtained in one year of running 
for the higher intensity option of the machine, 
providing $10^{21}$ useful $\mu^+$'s and $\mu^-$'s per year. 
A realistic detector of magnetized iron \cite{cdg} will be considered  
and detailed estimates of the corresponding expected backgrounds 
and efficiencies 
included in the analysis. Three reference detector distances are discussed: 
732 km, 3500 km and 7332 km. 

A preliminary version of this work was presented in \cite{ECFA}.

%
\section{ Fluxes and charged currents.}
%

\subsection{The Neutrino Factory}

One of the most encouraging outcomes of the recent neutrino factory 
workshop at Lyon \cite{lyon} 
was the convergence of the various machine designs existing previously, to an essentially 
unified design \cite{machine}, based on a muon accumulator with either a
triangular or a bow-tie shape. Both geometries permit two straight sections pointing in
different directions, allowing two different baselines.
 
It was also agreed in Lyon that the beam power on target should not exceed 4 MW. 
This, in turn, limits the production of muons to $10^{21}$ per year, out of which only 
20-25~\% are useful (that is, decay in the straight sections pointing towards the detectors). 
Agreement was also found upon other important parameters, namely the 
machine dual polarity (i.e, the ability to store $\mu^+$ and $\mu^-$, although
not simultaneously) and the maximum realistic energy to accelerate the
stored muons, which was fixed at 50 GeV.

Ultimate sensitivity to the neutrino mixing matrix parameters, in
particular to $\tetaot \,$ and $\delta$, require a data set 
as large as possible. 
The analysis presented in this paper assumes a total data set of $10^{21}$ 
useful $\mu^+$ decays and $10^{21}$ useful $\mu^-$ decays. 

\subsection{Number of events}

In the muon rest-frame, the distribution of muon antineutrinos (neutrinos) and 
electron neutrinos (antineutrinos) in the decay 
$\mu^\pm \raw e^\pm + \nu_e ( \bar\nu_e ) + \bar\nu_\mu ( \nu_\mu )$ 
is given by:
\be
\frac{d^2 N}{dx d\Omega}=\frac{1}{4\pi} [f_0 (x) \mp {\cal P}_\mu f_1 (x) \cos \theta] \, ,
\ee
where $E_\nu$ denotes the neutrino energy, $x=2E_{\nu}/m_{\mu}$ and 
${\cal P}_\mu$ is the average muon polarization along the beam directions. 
$\theta$ is the angle between the neutrino momentum vector and 
the muon spin direction and $m_\mu$ is the muon mass. 
The positron (electron) flux is identical to that for muon neutrinos 
(antineutrinos), when the electron mass is neglected. 
The functions $f_0$ and $f_1$ are given in Table~\ref{tab:tab1} \cite{gaisser}.
\begin{table}[t]
\centering
\begin{tabular}{||c|c|c||}
\hline\hline
    &  $f_0(x)$  & $f_1(x)$ \\
\hline\hline
$\nu_\mu, \rm{e}$ & $2 x^2 (3-2x)$ & $2 x^2 (1-2x)$ \\
\hline  
$\nu_e$ & $12 x^2 (1-x)$ & $12 x^2 (1-x)$ \\
\hline \hline
\end{tabular}
\caption{\it 
Flux functions in the muon rest-frame as in ref. \cite{gaisser}.}
\label{tab:tab1}
\end{table} 

In the laboratory frame, the neutrino fluxes, boosted along the muon momentum 
vector, are given by:
\bea
\frac{ d^2 N_{\bar \nu_\mu, \nu_\mu} }{ dy d\Omega} & = & 
   \frac{ 4 n_\mu }{ \pi L^2 m_\mu^6 } \,\  E_\mu^4 y^2 \, (1 - \beta \cos \varphi) 
   \,\, \left \{ \left [ 3 m_\mu^2 - 4 E_\mu^2 y \, (1 - \beta \cos \varphi) 
             \right ] \right . \nn \\
    & & \left. \mp \, {\cal P}_\mu 
   \left [ m_\mu^2 - 4 E_\mu^2 y \, (1 - \beta \cos \varphi)
             \right ] \right \} \, , \nn\\
\frac{ d^2 N_{\nu_e,  \bar\nu_e} }{ dy d\Omega} & = & 
   \frac{ 24 n_\mu }{ \pi L^2 m_\mu^6 } \,\, E_\mu^4 y^2 \, (1 - \beta \cos \varphi) 
   \,\, \left \{ \left [ m_\mu^2 - 2 E_\mu^2 y \, (1 - \beta \cos \varphi)
             \right ] \right . \nn \\
   & & \left. \mp \, {\cal P}_\mu 
   \left [ m_\mu^2 - 2 E_\mu^2 y \, (1 - \beta \cos \varphi)
             \right ] \right \} \, .
\label{fluxes}
\eea
Here, $\beta = \sqrt{1-m^2_\mu/E^2_\mu}$, $E_\mu$ is the parent muon energy, 
$y = E_\nu/E_\mu$, $n_\mu$ is the number of useful muons per year 
obtained from the 
storage ring and $L$ is the distance to the detector. $\varphi$ is the angle 
between the beam axis and the direction pointing towards the detector, assumed to be
located in the forward direction of the muon beam. The angular divergence will
be taken as constant, $\delta \varphi \sim 0.1$ mr.

Unlike traditional neutrino beams obtained from $\pi$ and $K$ decays, 
the fluxes in Eq.~(\ref{fluxes}), in the forward direction, present a 
leading quadratic dependence on $E_\nu$. As a consequence, 
the oscillation signal does not decrease with increasing $E_\mu$. 
In Appendix A we present our numerical results for the $\nu_e ( \bar \nu_e ) $ 
and $ \bar\nu_\mu ( \nu_\mu ) $ fluxes. 

The charged current neutrino and antineutrino interaction rates can be computed 
using the approximate expressions for the neutrino-nucleon cross sections with 
an isoscalar target,
\be
\sigma_{\nu N} \approx  0.67 \times 10^{-42} \times \frac{E_\nu}{{\rm GeV}} 
\times {\rm m^{2}}~, \qquad
\sigma_{\bar\nu N} \approx  0.34 \times 10^{-42} \times \frac{E_\nu}{{\rm GeV}} 
\times {\rm m^{2}}.
\ee
It follows that the number of charged current (CC) events at a neutrino factory
 scales cubically with energy.
In Appendix A we also include our numerical results for the  
rates of $e^\pm$ and $\mu^\mp$ production, from a $\mu^\mp$ beam.

\section{Oscillation Probabilities}
\label{sec:prob}

\subsection{In vacuum}
\label{vacuum}

Atmospheric or terrestrial experiments have an energy range such that 
$\Delta m^2 \, L / E_\nu \ll 1$ for the smaller ($\Delta m_{12}^2$)
but not necessarily for the larger ($\Delta m_{23}^2$) of these mass gaps.
Even then, solar and atmospheric transitions are not (provided $\tetaot \neq 0$) 
two separate two-generation oscillations. For $|\Delta m_{12}^2| \ll |\Delta m_{23}^2|$, 
neutrino oscillation probabilities at terrestrial distances are accurately described 
by only three parameters, $\tetatt$, $\Delta m_{23}^2 = \Delta m_{13}^2$ and $\tetaot$:
\begin{eqnarray}
P_{ \nu_e \nu_\mu ( \bar \nu_e \bar \nu_\mu ) } & = &  
s^2_{23} \, \sin^2 2 \tetaot \, \sin^2 \left ( \frac{ \delot \, L}{2} \right ) \, , \cr
P_{ \nu_e \nu_\tau (\bar \nu_e \bar \nu_\tau ) } & = &   
c^2_{23} \, \sin^2 2 \tetaot \, \sin^2 \left ( \frac{ \delot \, L}{2} \right ) \, ,\cr
P_{ \nu_\mu \nu_\tau ( \bar \nu_\mu \bar \nu_\tau ) } & = &  
c^4_{13} \, \sin^2 2 \tetatt \, \sin^2 \left ( \frac{ \delot \, L}{2} \right ) \, ,
\label{todasprobs}
\end{eqnarray}
where 
\be
\Delta_{ij} \equiv \frac{\Delta m^2_{ij} }{2 E_\nu} \, .
\ee
Eqs.~(\ref{todasprobs}) are a very good approximation when the 
solar parameters lie 
in the SMA-MSW or VO range. 
The present best fit value for $\theta_{13}$ 
is in the range $6^\circ$--$8^\circ$\cite{fogli, dgh2}, although it 
is compatible with zero within
 errors. 
Among the transitions in Eq.~(\ref{todasprobs})
 the channels $\nu_e \raw \nu_\mu, \nu_\tau$ 
have clearly the best sensitivity to a small $\tetaot$. 
Experimentally, the measurement of $\nu_e \raw \nu_\mu$ oscillations through 
the appearance of wrong-sign muons is far superior to that 
of $\nu_e \raw \nu_\tau$ 
oscillations through $\tau$ detection. In \cite{dgh}, it was shown that the 
sensitivity to $\tetaot$ of the former can improve the present 
limits, which are mainly set by Chooz \cite{chooz}, by at least 
two orders of magnitude. 

At the neutrino factory, precision measurements 
for $|\Delta m_{23}^2|$ and $\tetatt$ 
can also be performed. Measurements of $\tau$ appearance or $\mu$ disappearance may be 
competitive with wrong-sign $\mu$ signals for small values of $\tetaot$, 
because of the cosine dependence of the corresponding probabilities in Eqs.~(\ref{todasprobs}). 
We do not develop this topic further in the present work, see \cite{bgrw}.

In the LMA-MSW scenario, for which the effects of $\Delta m^2_{12}$
may be relevant at the neutrino factory,
a good and simple approximation for the $\nu_e \raw \nu_\mu$ transition probability 
is obtained by expanding to second order in the small parameters, 
$\tetaot, \Delta_{12} / \Delta_{13}$ and $\Delta_{12} L$:
\bea
P_{\nu_ e\nu_\mu ( \bar \nu_e \bar \nu_\mu ) } & = & 
s_{23}^2 \, \sin^2 2 \tetaot \, \sin^2 \left ( \frac{\delot \, L}{2} \right ) + 
c_{23}^2 \, \sin^2 2 \theta_{12} \, \sin^2 \left( \frac{ \Delta_{12} \, L}{2} \right ) \nn \\
& + & \tilde J \, \cos \left ( \pm \delta - \frac{ \delot \, L}{2} \right ) \;
\frac{ \Delta_{12} \, L}{2} \sin \left ( \frac{ \delot \, L}{2} \right ) \, , 
\label{vacexpand} 
\eea
where here and throughout the paper  the upper/lower sign in the formulae
 refers to 
neutrinos/antineutrinos,  and
\be
\tilde J \equiv c_{13} \, \sin 2 \theta_{12} \sin 2 \tetatt \sin 2 \tetaot 
\ee
is the usual combination of mixing angles appearing in the Jarlskog determinant. 
The first term in Eq.~(\ref{vacexpand}) is quadratic in $\sin \tetaot$, 
whereas the leading term in $\Delta_{12}$ is linear. 
The latter may then be significant or even dominant for very small values 
of $\tetaot$ \cite{romanino}, when $\Delta m^2_{12}$ and $\theta_{12}$ 
are in the range allowed by the LMA-MSW solution.

At ``short'' distances, such as 732 km, Eq.~(\ref{vacexpand}) can be 
further approximated by:
\be
P_{\nu_ e \nu_\mu (\bar \nu_e \bar \nu_\mu ) }  =   
  s_{23}^2 \, \sin^2 2 \tetaot \left ( \frac{\delot \, L}{2} \right )^2 
+ \tilde J \, \cos \delta \; \frac{\Delta_{12} \, L}{2} \frac{\delot \, L}{2}  
+ c_{23}^2 \sin^2 2 \theta_{12} \left ( \frac{\Delta_{12} \, L}{2} \right )^2.  
\label{probsl} 
\ee
The CP-odd term in the probability (i.e. the one proportional to $\sin \delta$) 
has dropped because it is of higher order in $\Delta_{ij} \, L$. 
The comparison of the two polarities is then not useful (except for doubling the statistics ), 
since the CP-conjugated channels measure the same probability. 
Furthermore, all terms in Eq.~(\ref{probsl}) have the same dependence on the neutrino energy 
and the baseline, and consequently is very hard to disentangle them. 
As a result we expect a large correlation between the parameters $\tetaot$, $\delta$, $\Delta_{12}$. 
Even though long baseline (LBL) reactor experiments \cite{reactors} will provide a measurement 
of $|\Delta m^2_{12}|$ if the LMA-MSW solution is at work, 
the parameters $\tetaot$ and $\delta$ 
will have to be determined simultaneously at the neutrino factory. 
It will then be necessary to go to longer baselines where the energy dependence 
of the different terms in Eq.~(\ref{vacexpand}) differs, and where
the comparison of the neutrino and antineutrino probabilities provides 
non-trivial information to separate $\delta$ from $\tetaot$.

It is uncertain  \cite{dfm} whether solar experiments will determine
 the sign of $\Delta m^2_{12}$ if it lies in the LMA-MSW range. 
If it remains unknown, it implies an ambiguity in the determination 
of $\delta$: 
to reverse the sign of $\Delta m^2_{12}$ in Eq.~(\ref{vacexpand}) is equivalent
to replacing $\delta$ by $\delta + \pi$. Notice, though, that whether
 there is CP-violation or not is independent of whether the phase in any parametrization is $\delta$ or $\delta+\pi$.

It is possible to construct a measurable CP-odd asymmetry, which in vacuum is proportional 
to $\sin \delta$. In refs.~\cite{dgh} and \cite{donini}, the authors considered 
the following integrated asymmetry (see also \cite{otherCP}):
\be
{\bar A}^{CP}_{e\mu} = \frac{ \{ N [\mu^-] / N_o [e^-] \}_+ 
                            - \{ N [\mu^+] / N_o [e^+] \}_-}{
                              \{ N [\mu^-] / N_o [e^-] \}_+ 
                            + \{ N [\mu^+] / N_o [e^+] \}_- } \, . 
\label{intasy}
\ee
The sign of the decaying muons is indicated by a subindex,
$N [\mu^+] ( N [\mu^-] ) $ are the measured number of wrong-sign muons, and 
$N_o [e^+] ( N_o [e^-] ) $ are the expected number of $\bar \nu_e (\nu_e)$ 
charged current interactions in the absence of oscillations. The significance 
of this asymmetry (i.e. the asymmetry divided by its statistical error) 
scales with the baseline and neutrino energy in the following way:
\be
\frac{ {\bar A}^{CP} }{ \delta {\bar A}^{CP} }  \propto   
   \sqrt{ E_\nu } \, \left | \, \sin \left ( \frac{ \deltt \, L}{2} \right ) \right | \, .
\ee
The best sensitivity to a non-zero CP-odd asymmetry is found at 
the maximum of the atmospheric oscillation. At the corresponding distance, however, 
matter effects are already important and should be taken into account, 
as we proceed to discuss in the next subsection. 

\subsection{In matter}

Of all neutrino species, only $\nu_e$ and $\bar \nu_e$ have  
charged-current elastic scattering amplitudes on electrons. 
This, as is well known, induces effective ``masses'' $\mu = \pm \, 2 \, E_\nu \, A$,
where the signs refer to $\nu_e$ and $\bar \nu_e$ and $A$ is the matter 
parameter,
\be
A = \sqrt{2} \, G_F \, n_e\,, 
\label{a}
\ee
with $n_e$ the ambient electron number density \cite{MSW}. 

Matter effects may be important if $A$ is comparable to, or bigger than, 
the quantity $\Delta_{ij}$ for some mass difference and neutrino energy, 
and if distances are large enough for the probabilities to be in the non-linear 
region of the oscillation. 

For the Earth's crust, with density $\rho \sim$ 2.8 g/cm$^3$ and roughly equal numbers of protons, 
neutrons and electrons, $A \sim 10^{-13}$ eV. 
The typical neutrino energies we are considering are tens of GeVs.
For instance, for $E_\nu=30$ GeV (the average $\bar\nu_e$ energy in the 
decay of $E_\mu=50$ GeV muons) $A = 1.1 \times 10^{-4}$ eV$^2$/GeV 
$\sim \deltt$. This means that matter effects will be important at long distances. 
Notice that at $L =$ 732 km and 3500 km the neutrino path remains in the 
Earth crust, whereas for 7332 km the deeper flight path meets a denser medium
  and 
$A = 1.5 \times 10^{-4}$ eV$^2$/GeV.
 
\subsubsection{Neglecting solar parameters: VO or SMA-MSW solutions}

Consider the case when $\Delta_{12}$ is negligible compared to 
$\deltt$, $L^{-1}$ and A. In the approximation of constant $n_e$,  
the transition probability in matter governing the appearance of wrong-sign
muons can then be read from \cite{Yasuda}:
\be
P_{\nu_e \nu_\mu ( \bar \nu_e \bar \nu_\mu) } \simeq  
s_{23}^2 \, \sin^2 2 \tetaot \left( \frac{ \delot }{ B_\mp } \right )^2 
\, \sin^2 \left( \frac{ B_\mp \, L}{2} \right )\ ,
\label{probmatt1}
\ee
where 
\be
B_\mp \equiv \sqrt{ \left [ \delot \, \cos 2 \tetaot \mp A \right ]^2
                   +\left [ \delot \, \sin 2 \tetaot \right ]^2 } \,.
\label{B}
\ee
 
For $A=0$, Eq.~(\ref{probmatt1}) reduces to the corresponding vacuum result:
the first line in Eq.~(\ref{todasprobs}). At short distances, that is for 
$B_\mp \, L/2$ sufficiently small, the sinus in Eq.~(\ref{probmatt1}) 
can be expanded and
\be
P_{\nu_e \nu_\mu ( \bar \nu_e \bar \nu_\mu ) } \sim 
s_{23}^2 \, \sin^2 2 \tetaot \,\left( \frac{ \delot \, L}{2} \right )^2 \, ,
\label{probmatt11}
\ee
which also coincides with the vacuum behaviour for small $\delot \, L/2$, 
even when $A \gg$ $\delot \cos 2 \tetaot$. Eq.~(\ref{probmatt11}) is a good
approximation 
up to $\sim$ 3000 km \cite{dgh}. Note that the leading dependence on $\tetaot$ is 
quadratic as in vacuum.

\begin{figure}[t]
\centering
\epsfig{file=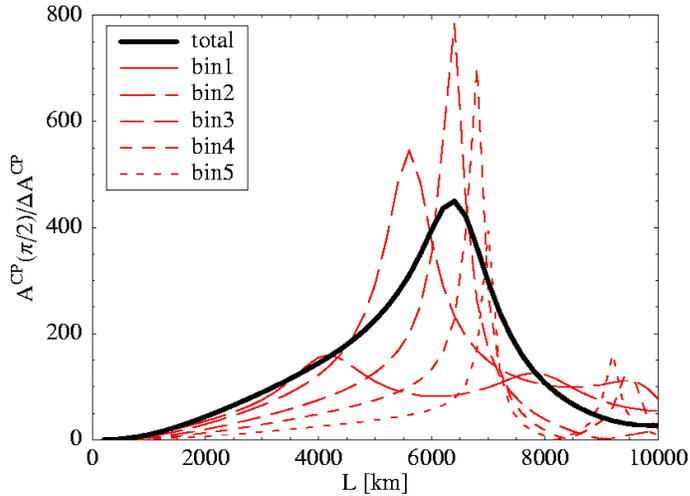, width=9cm} 
\caption{\it The signal-over-noise ratio of the CP-odd asymmetry of Eq.~(\ref{intasy}) 
as a function of the distance. The thick line corresponds to the integrated asymmetry, 
while the dashed lines correspond to the asymmetry computed in five energy bins 
of equal width $\Delta E_\nu = 10$ GeV. The neutrino mixing parameters correspond 
to the SMA-MSW solution to the solar anomaly:  
$\Delta m_{23}^2 = 2.8 \times 10^{-3}$ eV$^2$, $\Delta m_{12}^2 = 6 \times 10^{-6}$ eV$^2$, 
$\sin^2 2 \theta_{12} = 6 \times 10^{-3}$, $\tetatt = 45^\circ$, $\tetaot = 13^\circ$
and $\delta = 90^\circ$. The muon energy is $E_\mu =$ 50 GeV and the matter parameter $A$
is varied with the distance as in \cite{quigg}.} 
\label{fig:asbinsma}
\end{figure}

In contrast with the vacuum result, the probability in matter depends on 
the sign of $\Delta m^2_{13}$. It follows from Eqs.~(\ref{probmatt1}) and
(\ref{B})  
that a change on this sign  is equivalent to a CP transformation\footnote{Recall 
that in this approximation $\Delta m^2_{23} = \Delta m^2_{13}$.}, that is, 
interchanging the probability of neutrinos and antineutrinos. 
Thus matter effects induce by themselves a non-vanishing CP-odd asymmetry, 
and the best sensitivity to the sign is achieved when the sensititivity 
to the asymmetry of Eq.~(\ref{intasy}) is maximal. 
Fig.~\ref{fig:asbinsma} shows the significance of this asymmetry as a 
function of the baseline. The maximum sensitivity to the sign is thus 
expected at ${\cal O}(7000)$ km, although it is already very good at much shorter 
distances: notice the large number of standard deviations in the 
$y$ axis.

It is important to stress that having sizeable matter effects does not 
necessarily imply having sensitivity to the sign. 
For example if $A \gg \delot \cos 2 \tetaot$, the sensitivity is lost, 
even though matter effects are important at large distances. 
The optimal sensitivity occurs for $A \sim \delot \cos 2 \tetaot$, which is an 
energy dependent condition. In Fig.~\ref{fig:asbinsma} the asymmetry 
resulting from each of five energy bins of width $\Delta E_\nu = 10$ GeV 
is also shown: the asymmetries in different bins peak at different distances. 
This dependence suggests that using the information in energy bins can 
further improve the measurement of the sign of $\Delta m^2_{23}$.

\subsubsection{ With solar parameters: LMA-MSW solution}

In the LMA-MSW solar scenario, 
the effects of $\Delta m^2_{12}$ are not negligible 
over terrestrial distances, given the high intensity of the neutrino factory. 
The exact oscillation probabilities in matter when no mass difference is neglected have been derived analytically in \cite{zs}. However, the physical implications of 
the formulae in \cite{zs} are not easily derived. A convenient and 
precise approximation is obtained by expanding to second order 
in the following small parameters: 
$\tetaot$, $\Delta_{12}/\deltt$, $\Delta_{12}/A$ and $\Delta_{12} \, L$. 
The result is (details of the calculation can be found in appendix C):
\bea
P_{\nu_ e \nu_\mu ( \bar \nu_e \bar \nu_\mu ) } & = & 
s_{23}^2 \sin^2 2 \tetaot \, \left ( \frac{ \delot }{ \tilde B_\mp } \right )^2
   \, \sin^2 \left( \frac{ \tilde B_\mp \, L}{2} \right) \, + \, 
c_{23}^2 \sin^2 2 \theta_{12} \, \left( \frac{ \Delta_{12} }{A} \right )^2 
   \, \sin^2 \left( \frac{A \, L}{2} \right ) \nn \\
& + & \label{approxprob}
\tilde J \; \frac{ \Delta_{12} }{A} \, \frac{ \delot }{ \tilde B_\mp } 
   \, \sin \left( \frac{ A L}{2}\right) 
   \, \sin \left( \frac{\tilde B_{\mp} L}{2}\right) 
   \, \cos \left( \pm \delta - \frac{ \delot \, L}{2} \right ) \, , 
\label{hastaelmogno}
\eea
where $\tilde B_\mp \equiv |A \mp \delot|$. 
Once again, this expression reduces to the vacuum result, Eq.~(\ref{vacexpand}), 
in the limit $A \raw 0$. 
As already remarked in the previous subsection, a reversal of the sign of 
$\Delta_{12}$ in Eq.~(\ref{hastaelmogno}) can be simply traded by an shift of $\pi$ in $\delta$, and 
we will stick to positive $\Delta_{12}$ in the numerical exercises below.

We have numerically compared Eq.~(\ref{hastaelmogno}) with the exact formulae
 of \cite{zs}, in the range $1^\circ < \tetaot < 10^\circ$.
 Consider for instance the average energy $E_\nu= 30$ GeV and 
the following set of values: 
$\Delta m_{23}^2 = 2.8 \times 10^{-3}$ eV$^2$, $\theta_{12} = 22.5^\circ$,
$\tetatt = 45^\circ$ and $\delta=90^\circ$. 
%
For $\Delta m_{12}^2 = 1 \times 10^{-4}$ eV$^2$, 
the difference is $<$ 10 \%  ($<$ 20 \%) at $L \sim $ 3000 (7000) km. 
For $\Delta m_{12}^2 = 1 \times 10^{-5}$ eV$^2$, 
the error diminishes to $<$ 2.5 \% ($<$ 10 \%) at $L \sim $ 3000 (7000) km. 
Slightly better accuracy is obtained for $\delta=0^\circ$. 

As  before, 
matter effects in Eq.~(\ref{approxprob}) 
induce an asymmetry between neutrinos and antineutrinos oscillation 
probabilities even for 
vanishing $\delta$. 
For this reason the CP-odd asymmetry, Eq.~(\ref{intasy}), is not the most 
transparent observable to determine the optimal distance for measuring 
 $\delta$. 
A better way of addressing the issue
is to ask at what distance the significance of the terms 
which depend on $\delta$ is maximal.

\begin{figure}[t]
\begin{center}
\begin{tabular}{ll}
\hskip -0.5cm
\epsfig{file=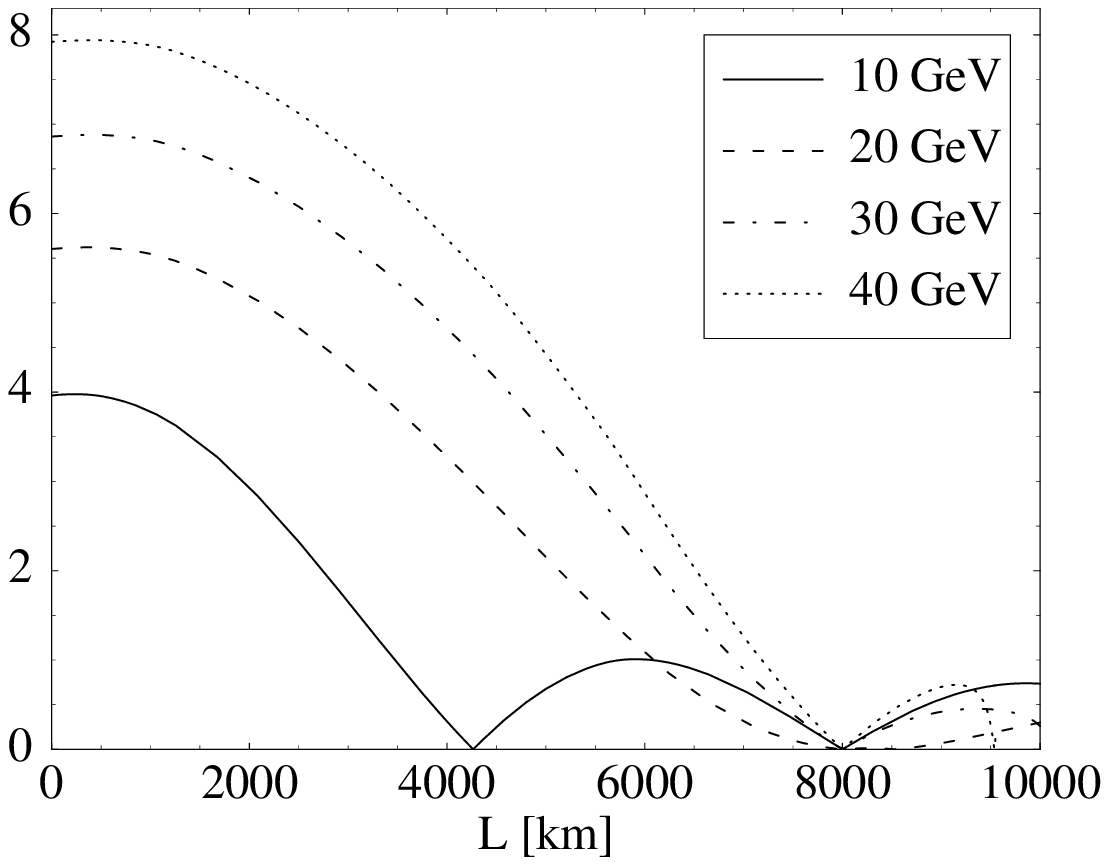, width=8.1cm} &
\hskip -0.5cm
\epsfig{file=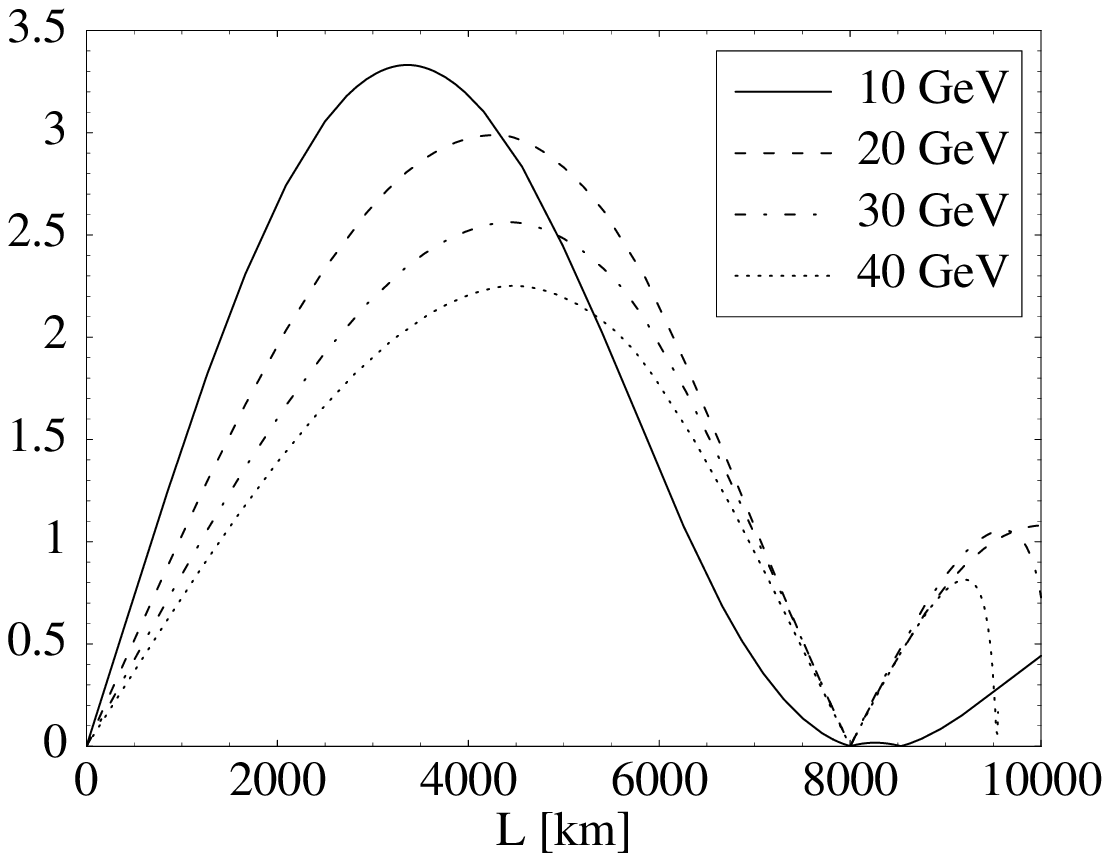, width=8.3cm} \\
\hskip 3.2truecm
{\small (a)}            &
\hskip 3.3truecm
{\small (b)}
\end{tabular}
\end{center}
\caption{\it a) Scaling with $L$ of the ratio of the number of 
wrong-sign muons $\mu^-$ induced solely by $P_2$ in Eq.~(\ref{pis}) 
to the statistical error in the total number of wrong-sign muons, for 
$\delta = 0^\circ$; b) the same for wrong-sign muons $\mu^-$ induced by $P_3$, 
for $\delta = 90^\circ$. 
The normalization of the $y$ axis is arbitrary. 
The oscillation parameters are: $\Delta m_{23}^2 = 2.8 \times 10^{-3}$ eV$^2$, 
$\Delta m_{12}^2=1 \times 10^{-4}$ eV$^2$, $\theta_{12} = 22.5^\circ$, 
$\tetatt = 45^\circ$ and $\tetaot = 8^\circ$. }
\label{cossin}
\end{figure}

For $\tetaot \ge 1^\circ$, the dominant contributions in Eq.~(\ref{approxprob})
 are:
\bea
P_1 & \equiv & s_{23}^2 \sin^2 2 \tetaot \left ( \frac{ \delot }{ \tilde B_\mp} \right )^2 
\, \sin^2 \left( \frac{ \tilde B_\mp \, L}{2} \right ) \, , \nn \\
P_2 & \equiv & \tilde J \; \frac{ \Delta_{12} }{A} \frac{ \delot }{ \tilde B_\mp} 
\, \cos \delta \cos \left( \frac{ \delot \, L}{2} \right ) \; 
               \sin \left( \frac{ A \, L}{2} \right ) 
               \sin \left( \frac{ \tilde B_\mp \, L}{2} \right ) \, , \nn \\ 
P_3 & \equiv & \pm \tilde J \; \frac{ \Delta_{12} }{A} \frac{ \delot }{ \tilde B_\mp} 
\, \sin \delta \sin \left ( \frac{ \delot \, L}{2} \right ) \; 
               \sin \left ( \frac{ A \, L}{2} \right ) 
               \sin \left ( \frac{ \tilde B_\mp \, L}{2} \right ) \, . 
\label{pis}
\eea

In Fig.~\ref{cossin} we show the significance of the $\delta$-dependent terms,
$P_2$ and $P_3$, as function of $L$. The significance is defined as 
the fraction 
of wrong-sign muons--for positive muons in the beam-- resulting from 
 a given term, 
over the statistical error in the measurement 
of the total number of wrong-sign muons. Results for several values of 
the neutrino energy are depicted. 

\begin{figure}[t]
\centering
\begin{tabular}{ll}
\hskip -1cm
\epsfig{file=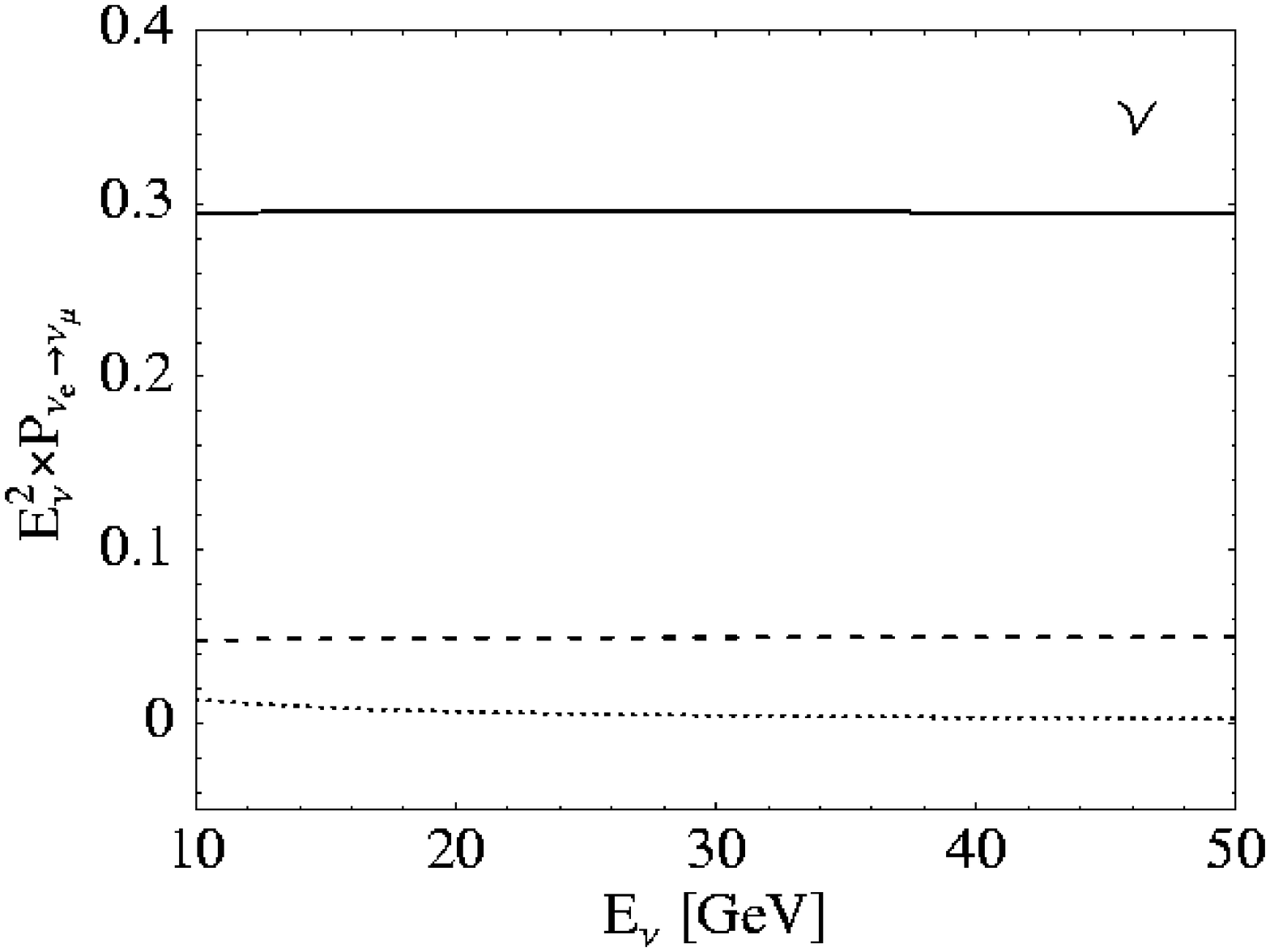, width=8cm} &
\epsfig{file=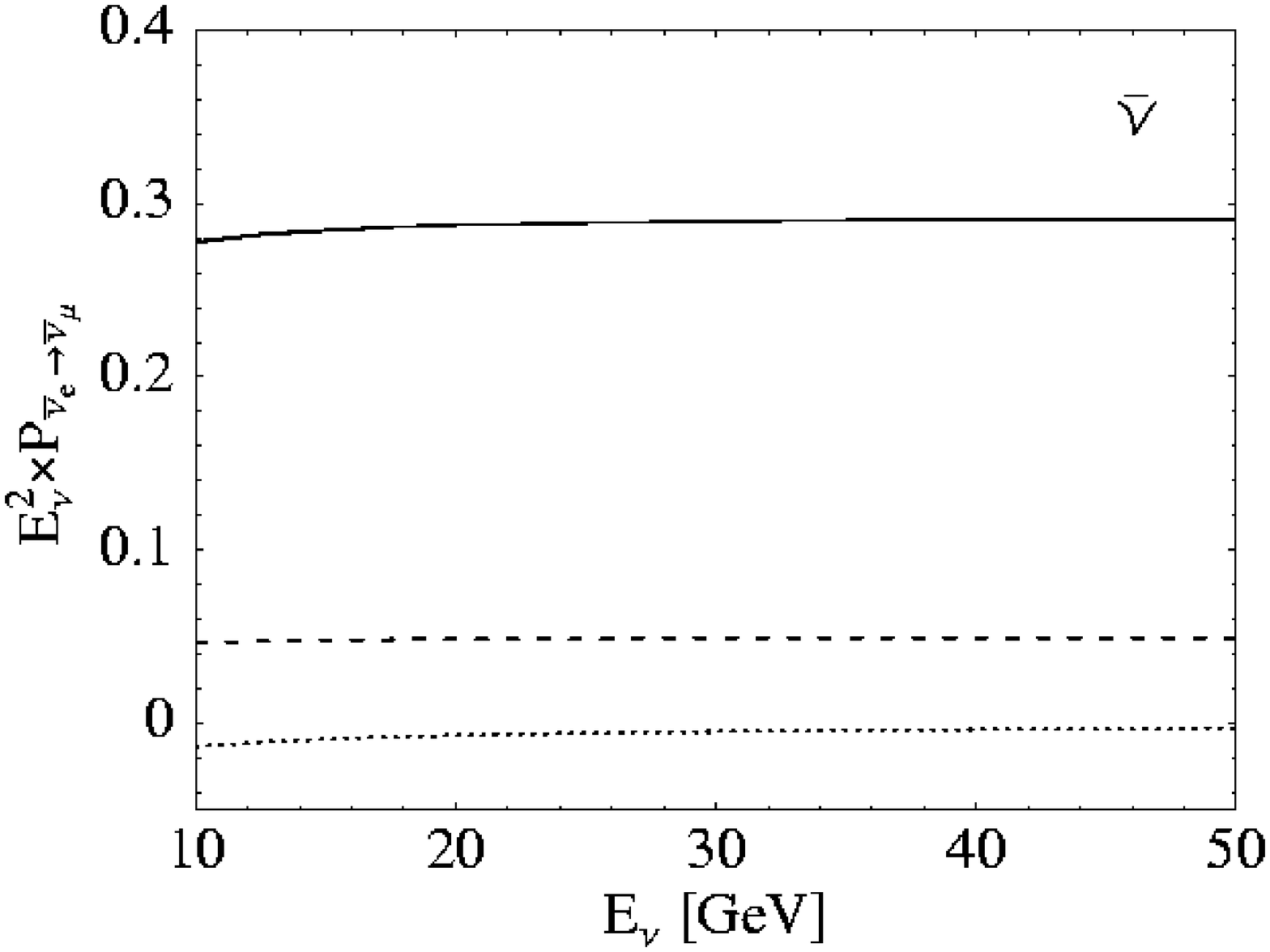, width=8cm} \\
\hskip 3.2truecm
{\small (a)}            &
\hskip 4.2truecm
{\small (b)}
\end{tabular}
\caption{\it Neutrino energy dependence of the different terms in Eqs.~(\ref{pis}): 
$E^2_\nu \, P_1 $ (solid line), $E^2_\nu \, P_2$ $\delta =0^\circ$ (dashed line) for  and 
$E^2_\nu \, P_3$ for $\delta = 90^\circ$ (dotted line) , at $ L = $ 732 km, 
for neutrinos (a) and antineutrinos (b). The parameters correspond to the LMA-MSW solution: 
$\Delta m_{23}^2 = 2.8 \times 10^{-3}$ eV$^2$, $\Delta m_{12}^2 = 1 \times 10^{-4}$ eV$^2$, 
$\theta_{12} = 22.5^\circ, \tetatt = 45^\circ$ and $\tetaot = 8^\circ$.} 
\label{corr1}
\end{figure}

The term in $\cos \delta$, $P_2$, is 
more significant than $P_3$ at short distances.
Unfortunately, this sensitivity to $\delta$ through $P_2$ 
is fake, because at short distances there is no way to separate 
$P_2$ from the leading term, $P_1$: 
they have similar energy dependence and do not differ in the two polarities, as illustrated 
in Fig.~\ref{corr1}. In order words, a change in $\delta$ can be compensated by a change in 
$\tetaot$ to keep the total probability unchanged. 
\begin{figure}[t]
\centering
\begin{tabular}{ll}
\hskip -1cm
\epsfig{file=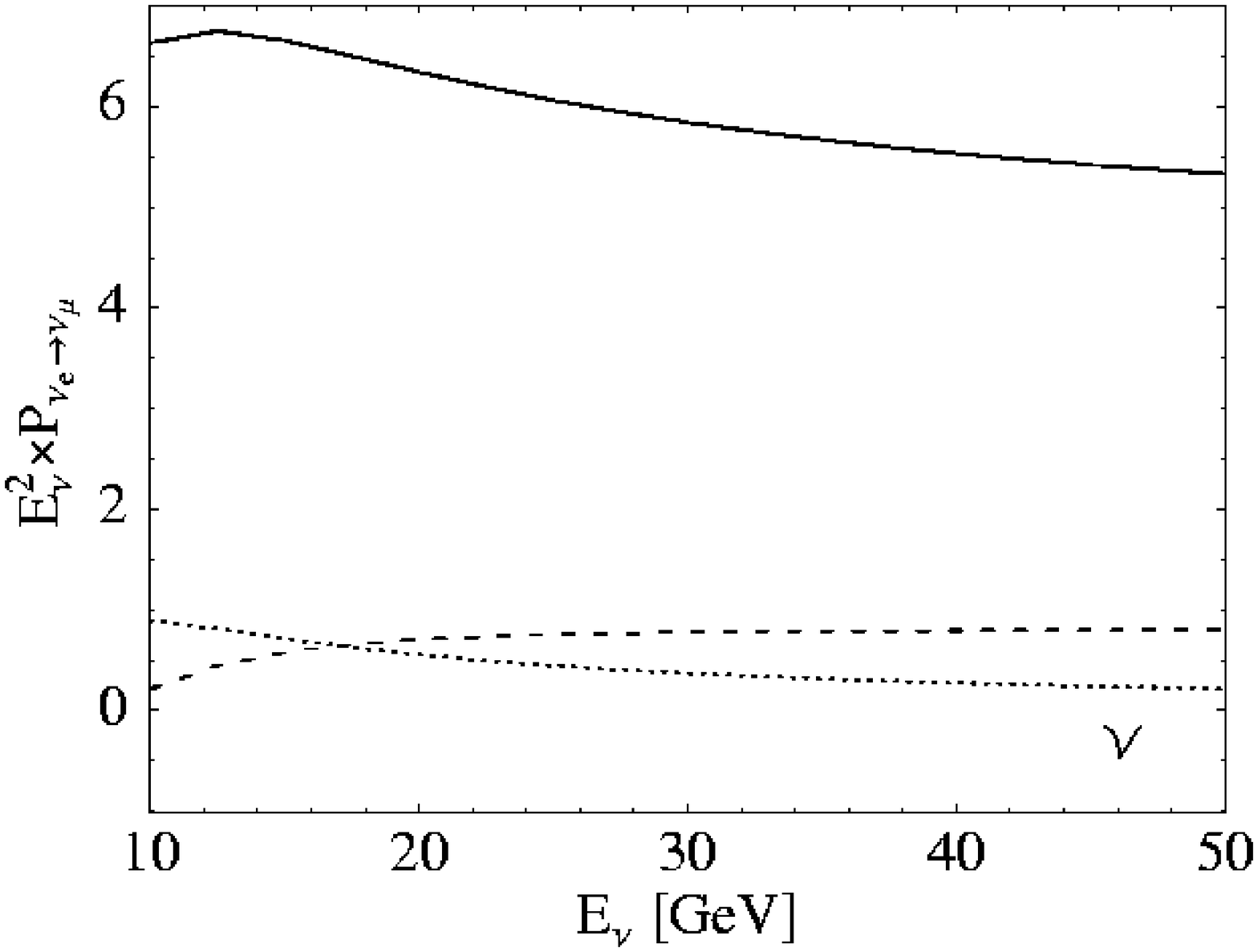, width=8cm} &
\epsfig{file=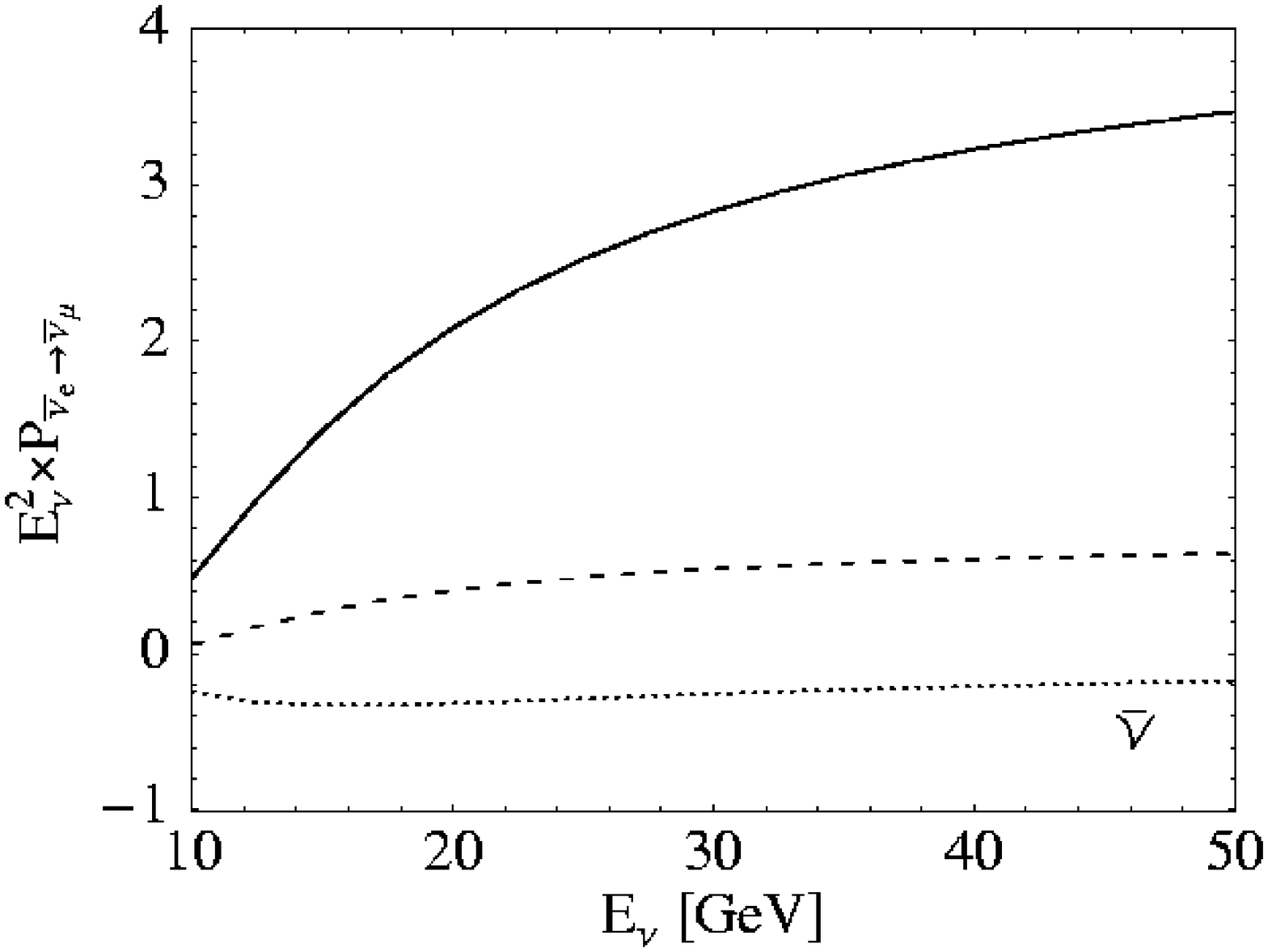, width=8.2cm} \\
\hskip 3.2truecm
{\small (a)}            &
\hskip 4.2truecm
{\small (b)}
\end{tabular}
\caption{\it The same as in Fig.~\ref{corr1} but for $L = $ 3500 km.}
\label{corr2}
\end{figure}

\begin{figure}[t]
\begin{center}
\epsfig{file=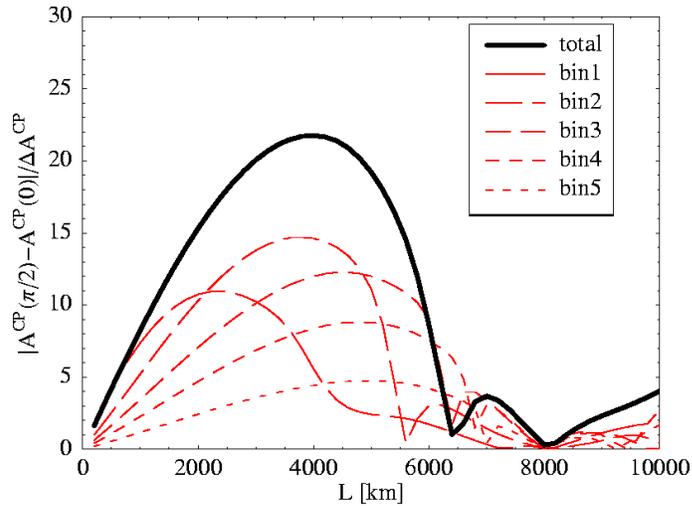, width=9cm}
\end{center}
\caption{\it 
The signal-over-noise ratio of the CP-odd asymmetry of Eq.~(\ref{intasy}) 
as a function of the distance, after subtracting the fake matter induced asymmetry. 
The thick line corresponds to the integrated asymmetry, while the dashed lines correspond 
to the asymmetry computed in five energy bins of equal width $\Delta E_\nu = 10$ GeV. 
The neutrino mixing parameters correspond to the LMA-MSW solution to the solar anomaly:  
$\Delta m_{23}^2 = 2.8 \times 10^{-3}$ eV$^2$, $\Delta m_{12}^2 = 1 \times 10^{-4}$ eV$^2$, 
$\theta_{12} = 22.5^\circ$, $\tetatt = 45^\circ$, $\tetaot = 13^\circ$ and $\delta = 90^\circ$. 
The muon energy is $E_\mu =$ 50 GeV and 
the matter parameter $A$ is varied with the distance as in \cite{quigg}.} 
\label{fig:aslma}
\end{figure}

At 3500 km the situation is very different, though: 
as Fig.~\ref{corr2} illustrates, 
the energy dependence of the three terms is quite different 
and furthermore $P_3$ 
--which changes sign with the beam polarity-- is considerably larger. 
The comparison of the number of wrong-sign muons detected running in 
the two polarities 
and the binning in energy of the signal are thus strong analysis tools to 
disentangle $\tetaot$ and $\delta$ at baselines around 3500 km.  
Notice that this optimal distance is 
in nice agreement with the previous studies based on the significance 
of integrated asymmetries \cite{dgh,donini,ourprocs}, updated  in Fig.~\ref{fig:aslma} 
for the present set-up.

The summary of this long discussion is that baselines 
much larger than 732 km
are needed, for the following reasons:
\begin{itemize} 
\item 
In the SMA-MSW or VO scenarios, the sign of $\Delta m^2_{23}$ can only be 
determined for distances such that matter effects are sizeable and the CP asymmetries 
they induce measurable. This happens at $L= {\cal O}$ (3000 km) or larger. 
\item 
In  the LMA-MSW scenario, there is a strong correlation between $\tetaot$ 
and $\delta$ at short distances. It is necessary to go far so that
the terms most sensitive to them show a different energy dependence,
 and the signals in the CP-conjugate channels differ sizeably, allowing 
the simultaneous measurement of both parameters.

\end{itemize}

These expectations will be sustained by a detailed energy analysis in 
the following sections.

\section{Detection of wrong sign muons}

\subsection{A Large Magnetized Calorimeter}

\begin{figure}
\begin{center}
\mbox{\epsfig{file=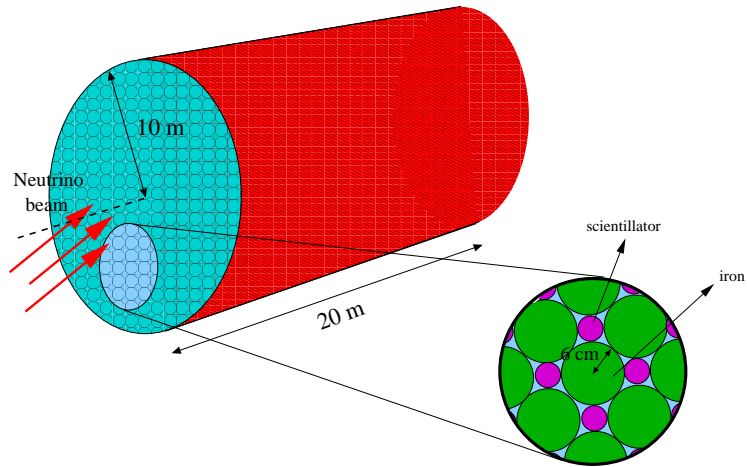,width=10cm}}
\end{center}
\caption{\it Sketch of the Large Calorimeter for the Neutrino Factory.} 
\label{fig:detector}
\end{figure}

For the present study we consider
a Large Magnetized Iron Calorimeter such as the one proposed in \cite{cdg}.
The apparatus, shown in Fig.~\ref{fig:detector} is a huge cylinder, 
of $10$ m radius and $20$ m length. It is made of $6$ cm thick iron rods 
intersped with $2$ cm thick scintillators segmented along their length. 
Its fiducial mass is $40$ kT. A superconducting coil generates a 
solenoidal magnetic field of 1 T.

The detector axis is oriented to form a few degrees with the direction of 
the neutrino beam. Thus, a neutrino crossing  the detector
sees a sandwich of iron and scintillator. The coordinates of the track 
transverse to the cylinder axis are measured 
from the location of the scintillator  rods, 
while the longitudinal coordinate 
is measured from their segmentation. As discussed in
\cite{cdg}, the performance of the device would be similar to
the one expected for the MINOS detector \cite{minos}.

Neutrino interactions in such a detector have a clear signature. For example, a
\numu\ charged current (CC) event is characterized by a muon, 
which is seen as a penetrating
long track, plus a shower resulting from the interactions of the hadrons
in the event. Fitting the muon track determines its charge and momentum, and
a calorimetric measurement allows the determination of the hadronic momentum 
vector. 
In contrast, for a \nue\ CC event the electromagnetic shower due to the prompt
electron cannot be disentangled from the hadronic shower in an event-by-event
basis, and therefore, the event looks similar to a neutral current (NC), which
is characterized by having no penetrating track.

A realistic simulation of the response of an iron calorimeter must be able
to compute:  1) whether the primary muon 
characterizing a CC event was identified or not, 2) whether non prompt
muons arising from the decays of hadrons were missidentified as primary muons and 3) the muon and hadronic momentum vectors.

To address the above points we have written a Monte 
Carlo simulation based in the GEANT 3 package \cite{geant}. The apparatus is 
simulated with the correct geometry, and neutrino 
interactions  are generated at random points in the fiducial 
volume. Then, every  particle produced
in the interaction is followed until it decays, exits the 
detector or undergoes a nuclear interaction. 

When the muon track is very short, it cannot be disentangled
from the other tracks in the hadronic jets. The
peak of the hadron shower occurs at about 10 cm from
the interaction vertex and 
essentially no hadronic activity remains at 100 cm from it. We thus  impose the
conservative criteria that, in order to be reconstructed, a muon
track must be longer than 100 cm. About 99.2 \% of all \numu\ CC events
at 50 GeV produce primary muons that satisfy this condition.

All muons, either primary or arising from the decay of hadrons, 
are tracked through the entire volume, and a hit is recorded each time
a scintillator rod is crossed. The tracks can then be fitted to obtain the muon momentum. However, once the distance travelled by the muon is known, it 
is  more efficient to use a simple smearing, which takes  into account 
correctly both the effect of detector resolution and the multiple scattering. 

\begin{figure}
\begin{center}
\mbox{\epsfig{file=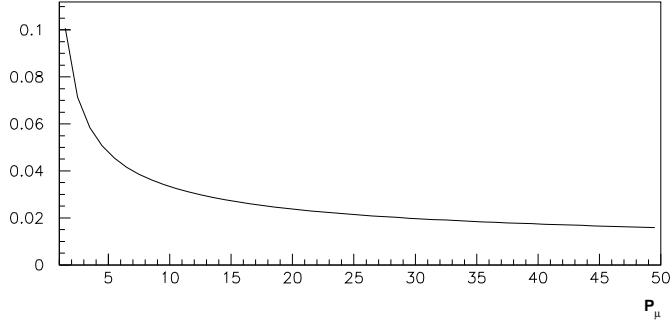,width=10cm}}
\end{center}
\caption{\it Relative error 
in the measurement of the muon momentum as a function
of the muon momentum.}
\label{fig:mome}
\end{figure}

As an illustration of the smearing procedure,  
the relative error in the measurement of the muon momentum as a function
of the muon momentum is shown  
in Fig. \ref{fig:mome}. The resolution decreases 
rapidly for low momentum muons (for
$P_\mu < 5$ GeV, a much better resolution
would be obtained from the measurement of the muon range), and improves
rather smoothly for large momentum ($\delta P_\mu/P_\mu \sim 4 \%$ for
$P_\mu \sim 7~$GeV, while  $\delta P_\mu/P_\mu \sim 3 \%$ for
$P_\mu \sim 50~$GeV). The fact that $\delta P_\mu/P_\mu$ is almost constant
for large momentum is due to the dominance of the
multiple scattering term in the resolution.

Hadrons, in contrast, are followed until they decay or undergo 
a nuclear interaction. In the latter case their momenta are  
added to the hadronic energy vector. Finally, both the magnitude and direction
of the  hadronic vector was smeared to account for the resolution of
the detector. For details we refer to \cite{cdg}.

A large sample
of neutrino interactions, corresponding to $10^7$ \numubar\ 
charged and neutral currents, $5\times 10^6$ \nue\ charged 
and neutral currents and the same data samples for the opposite polarity 
were analyzed for this study.

\subsection{Search for wrong-sign muon events}
  
We consider first the (\numubar\ , \nue\ ) neutrino beams originating from a
$\mu^+$ beam of $E_\mu=50$ GeV (the dependence of the backgrounds with the
energy of the muon beam was discussed in \cite{cdg}, where it was shown that 
optimal performance is obtained at the highest possible energy).
The bulk of the events in the detector are
\numubar\ charged currents, signaled by the presence of a positive primary
muon in the event, \numubar\ and \nue\ neutral currents, which are events
with no primary leptons, and \nue\ charged currents, for which we assume that 
the primary electron is not identified. 
On top of those events, one searches for
wrong sign $\mu^-$ arising from the \numu\ produced via the oscillation 
$\nu_e \rightarrow \nu_\mu$. 
Table \ref{tab:mupsample} shows the number of interactions corresponding
to a total of $10^{21}$ useful $\mu^+$ decays and a 40 kT detector at our
reference baselines. For illustration, we will consider the oscillation 
parameters for the signal in the LMA-MSW scenario: $\Delta m^2_{23} = 4\times 10^{-3} $eV$^2$, $\Delta m^2_{12} = 10^{-4}$ eV$^2$, $\theta_{13}=13^\circ$, $\theta_{12} = 22.5^\circ$ and $\theta_{23} = 45^\circ$, as in the tables of Appendix B.
 Notice however that our results for the fractional background and efficiencies should be rather insensitive to the particular choice of parameters.  
\begin{table}[bhtp]
\begin{center}
\begin{tabular}{|c|c|c|c|c|} \hline
  Baseline (km) & \numubar\ CC   & \nue\ CC 
& \numubar\ + \nue\ NC & 
\numu (signal) 
\\ \hline \hline
732 & $3.5 \times 10^7$ & $5.9 \times 10^7$  & $3.1 \times 10^7$ & 
$1.1 \times 10^5$ \\
3500 & $1.5 \times 10^6$ & $2.6 \times 10^6$  & $1.3 \times 10^6$ & 
$1.0 \times 10^5$ \\
7332 & $3.5 \times 10^5$ & $5.9 \times 10^5$  & $3.0 \times 10^5$ & 
$3.8 \times 10^4$ \\
\hline
\end{tabular}
\end{center}
\caption{\it Data samples expected in a 40 kT detector for
$10^{21}$ useful $\mu^+$ decays. $\nu_e \rightarrow \nu_\mu$ 
oscillations with parameters as in Appendix B.}
\label{tab:mupsample}
\end{table}

The potential backgrounds to the wrong sign $\mu^-$ events signaling the
presence of oscillations are:
\begin{enumerate}
\item \numubar\ CC events in which the positive muon is not
detected, and a secondary negative muon arising from the decay of
$\pi^-, K^-$ and $D^-$ hadrons fakes the signal. The most
dangerous events are those with $D^- \rightarrow \mu^-$, which yield
an energetic muon with a spectrum similar to the signal. 
\item \nue\ CC events, for which it is assumed that the primary electron
is never detected. Charm production is not relevant for this type of events
since the charmed hadrons in the hadronic jet are
predominantly positive. Instead, fake $\mu^-$ arise 
from the decay of negative pions and kaons in the hadronic jet.
\item \numubar\ and \nue\ NC events. Fake $\mu^-$ arise in this case
also predominantly from the decay of negative pions and kaons, since
charm production is suppressed with
respect to the case of CC.
\end{enumerate} 

\begin{figure}
 \begin{center}
 \mbox{\epsfig{file=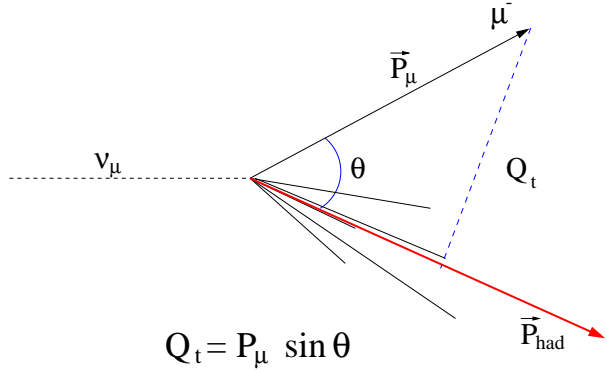,width=8cm}}
 \end{center}
\caption{\it Definition of the kinematical variables used in this study.}
\label{fig:qt}
\end{figure}

At first sight these backgrounds seem impressive. 
Fortunately, 
simple
kinematical cuts can suppress them very efficiently. 
One exploits the fact that for signal events the
$\mu^-$ candidate is harder and more isolated from the hadronic
jet than for background events. We thus perform a simple analysis
based in two variables, namely: the momentum of the muon, $P_\mu$, and 
a variable
measuring the isolation of the muon, $Q_t = P_\mu \sin \theta$
(see Fig. \ref{fig:qt}).

\begin{figure}
\begin{center}
\numubar Charged Currents
\mbox{
\epsfig{file=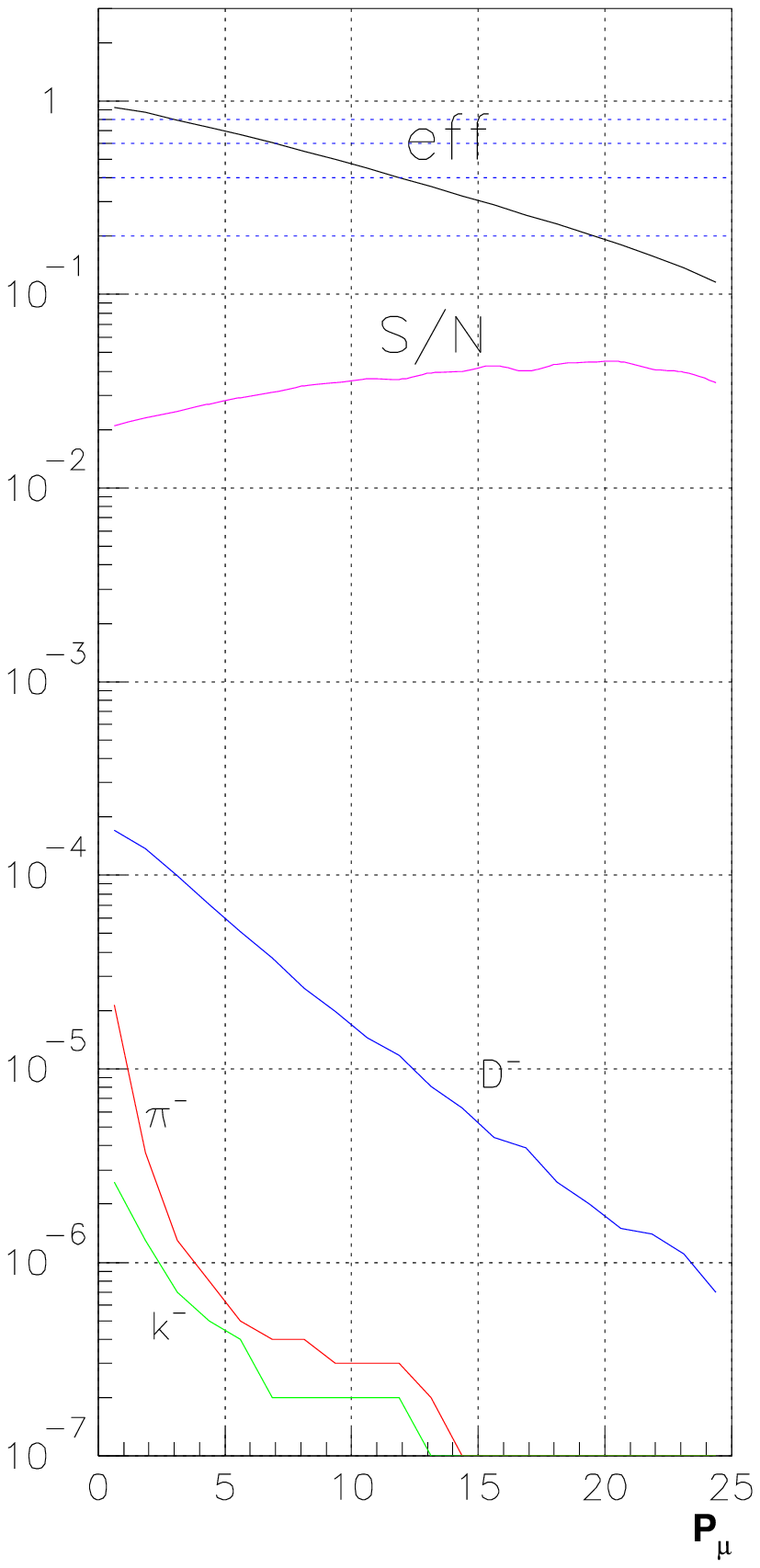,height=7cm,width=7cm}
\epsfig{file=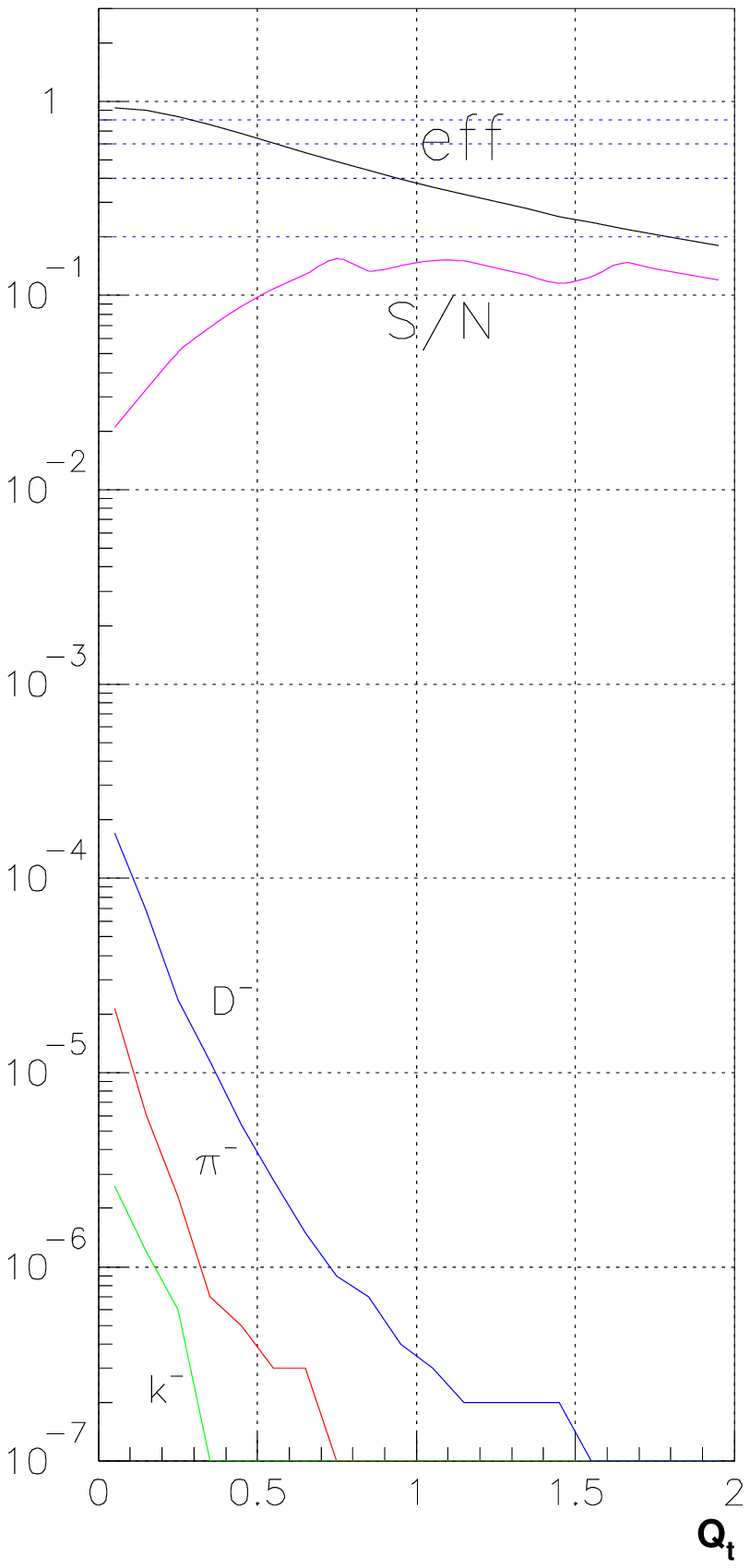,height=7cm,width=7cm}
}
\\
\numubar Neutral Currents
\mbox{
\epsfig{file=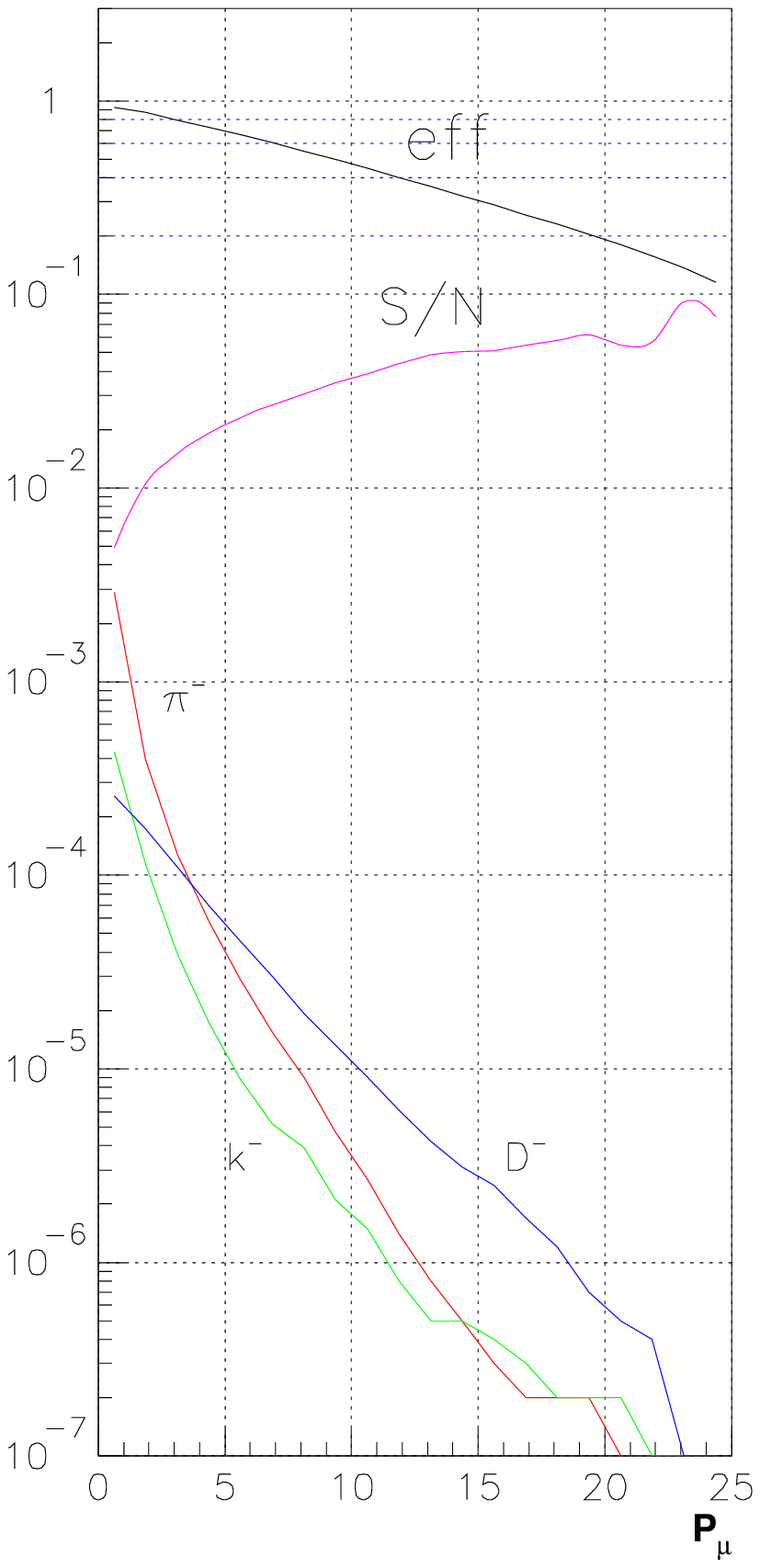,height=7cm,width=7cm} 
\epsfig{file=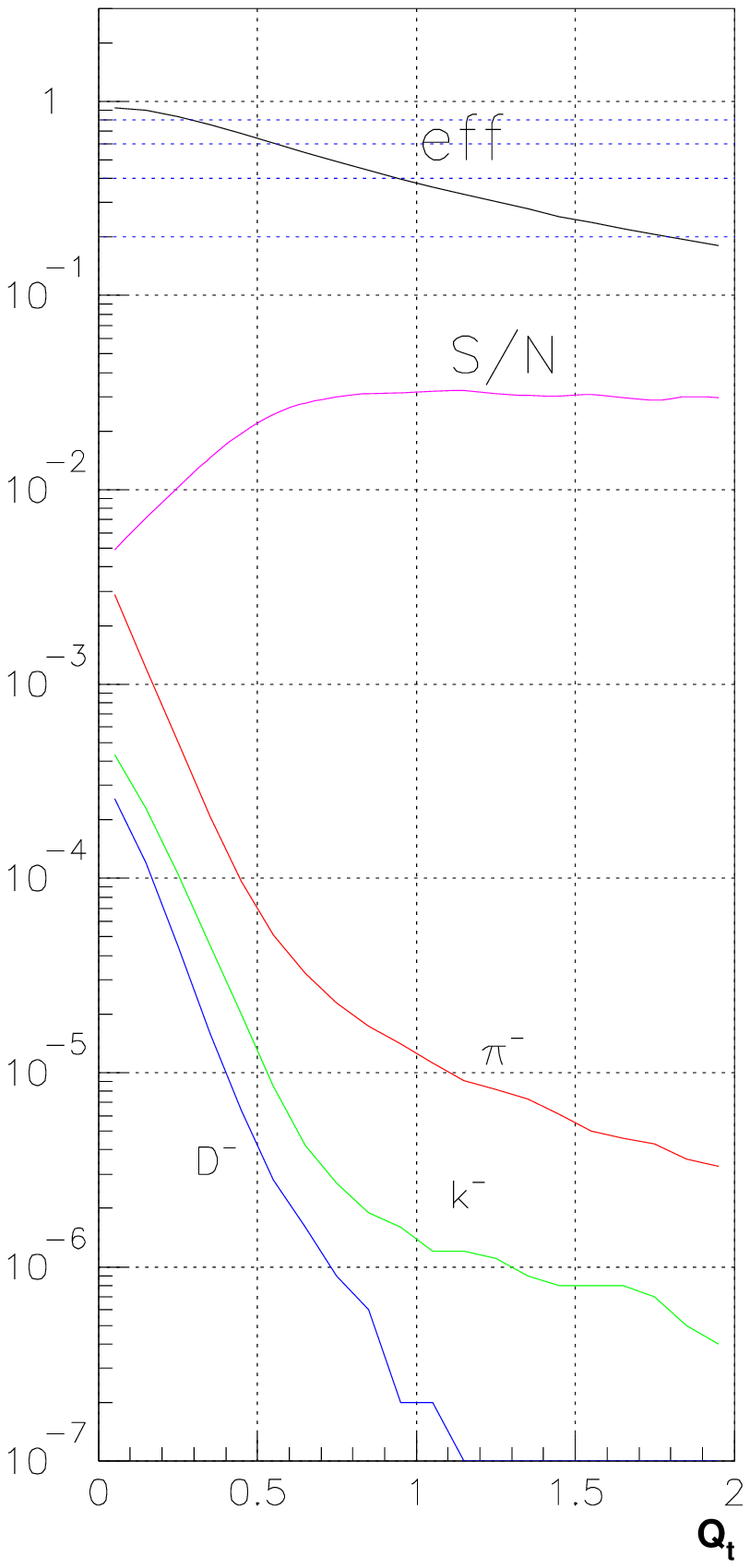,height=7cm,width=7cm}  
}    
\end{center}
\caption{\it Signal and backgrounds for \numubar\ CC and NC events.}
\label{fig:ana1}
\end{figure}

To illustrate the rejection power of this analysis, 
Fig. \ref{fig:ana1} 
shows the efficiency for signal detection and
the fractional backgrounds as a function of $P_\mu$ and $Q_t$ for
\numubar\ charged and neutral currents. 
Also shown is the signal-to-noise ratio, $S/N$, defined as the ratio of 
the signal selection efficiency and the error in the subtraction of the
 number of 
background events that pass the cuts, $N_b$. The error is taken to be Gaussian for large $N_b$ ($\sim \sqrt{N_b}$) and Poisson otherwise.
Notice that 
charm production is the dominant background for 
\numubar\ CC, while $\pi$ decay dominates the
NC backgrounds. 

Inspection of Fig. 
\ref{fig:ana1} shows that the $S/N$ is rather flat
for $P_\mu > 5~$GeV and $Q_t >0.5~$GeV. 
Cutting at
$P_\mu > 7.5~$GeV, $Q_t > 1.0~$GeV, maximizes $S/N$. However,
given the flatness of the $S/N$ ratio, one can vary these values 
generously with very little difference in the final results.
Table \ref{tab:mueff} shows the fractional background contamination and the 
signal efficiency, while 
Table \ref{tab:muplus} shows the number of background and signal events that
pass the cuts for the data sample discussed above. Notice that
the residual backgrounds are quite sizeable at $L = 732 ~$km, 
small at $L = 3500 ~$km and negligible at $L = 7332~$km.
It is possible to optimize the analysis for very long baseline
in order to achieve higher efficiency. However, we have chosen
to use the same set of cuts for the three baselines.
\begin{table}[bhtp]
\begin{center}
\begin{tabular}{|c|c|c|c|} \hline
    \numubar\ CC   & \nue\ CC 
& \numubar + \nue NC & 
\numu (signal) 
\\ \hline \hline
$1.0 \times 10^{-7}$ & $5.0 \times 10^{-7}$  & $1.0 \times 10^{-6}$ & 
$3 \times 10^{-1}$ \\
\hline
\end{tabular}
\end{center}
\caption{\it Fractional backgrounds and signal selection efficiency for 
the wrong sign muon search with $\mu^+$ decays.}
\label{tab:mueff}
\end{table}

\begin{table}[bhtp]
\begin{center}
\begin{tabular}{|c|c|c|c|c|} \hline
  Baseline (km) & \numubar\ CC   & \nue\ CC 
& \numubar + \nue NC & 
\numu (signal) 
\\ \hline \hline
732 & $3.5 $ & $30$  & $31$ & 
$3.3 \times 10^4$ \\
3500 & $0.1$ & $1.2$  & $1.2$ & 
$3 \times 10^4$ \\
7332 & $<0.1$ & $ <0.2 $  & $ <0.2$ & 
$1.3 \times 10^4$ \\
\hline
\end{tabular}
\end{center}
\caption{\it Events surviving the cuts in a 40 kT detector for
$10^{21}$ useful $\mu^+$ decays. Oscillation parameters as in 
Appendix B.}
\label{tab:muplus}
\end{table}

A potential source of fake wrong sign muons not discussed here is
that due to a wrong measurement of the charge. In \cite{cdg}
it was estimated that, for $E_\mu = 50 ~$GeV, 
this background could be reduced to a
very small level, of the order of
$10^{-6}$ or less. We have not included this source of background in 
the analysis.

\begin{figure}
\begin{center}
\numu Charged Currents
\mbox{
\epsfig{file=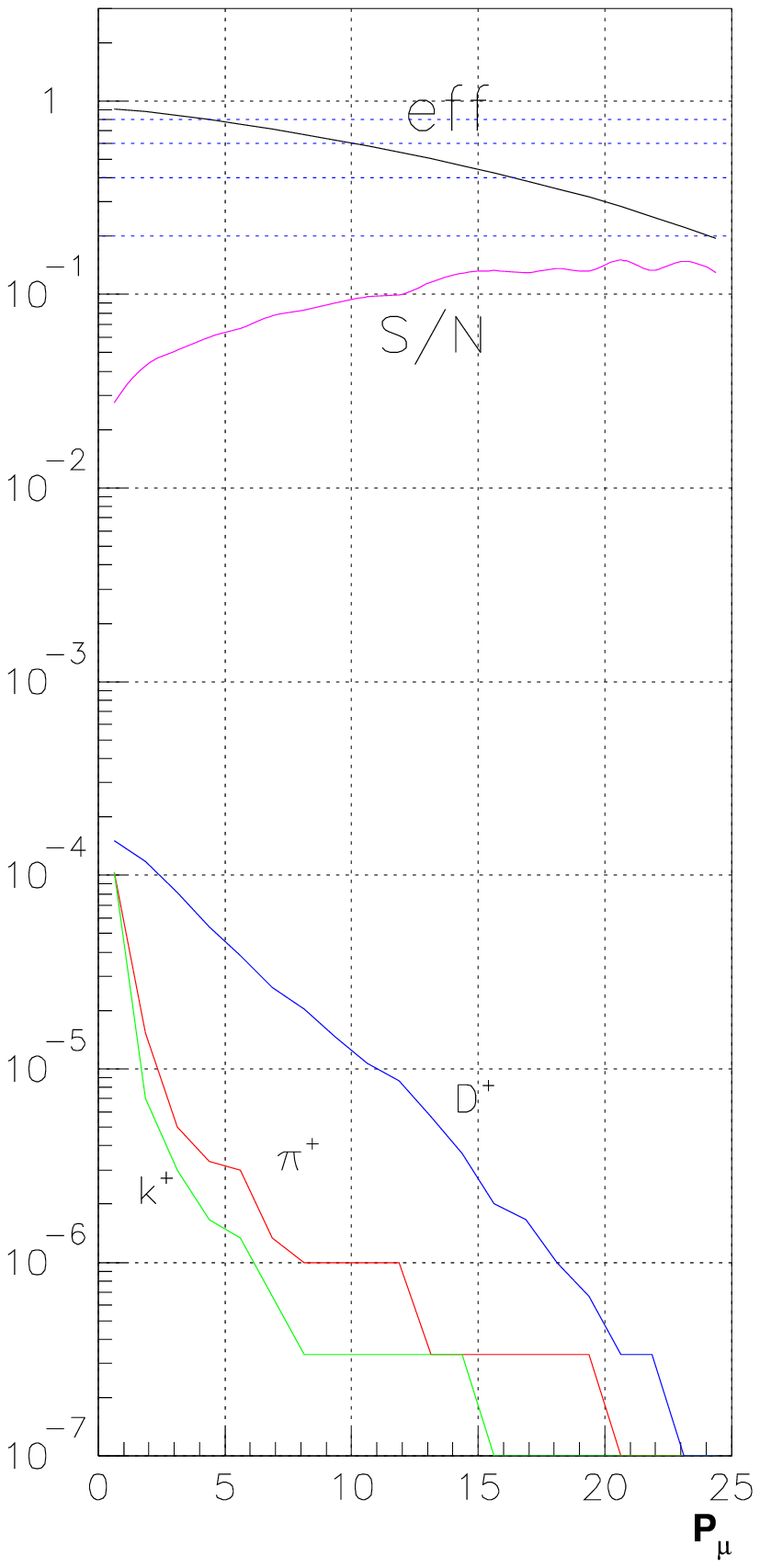,height=7cm,width=7cm}
\epsfig{file=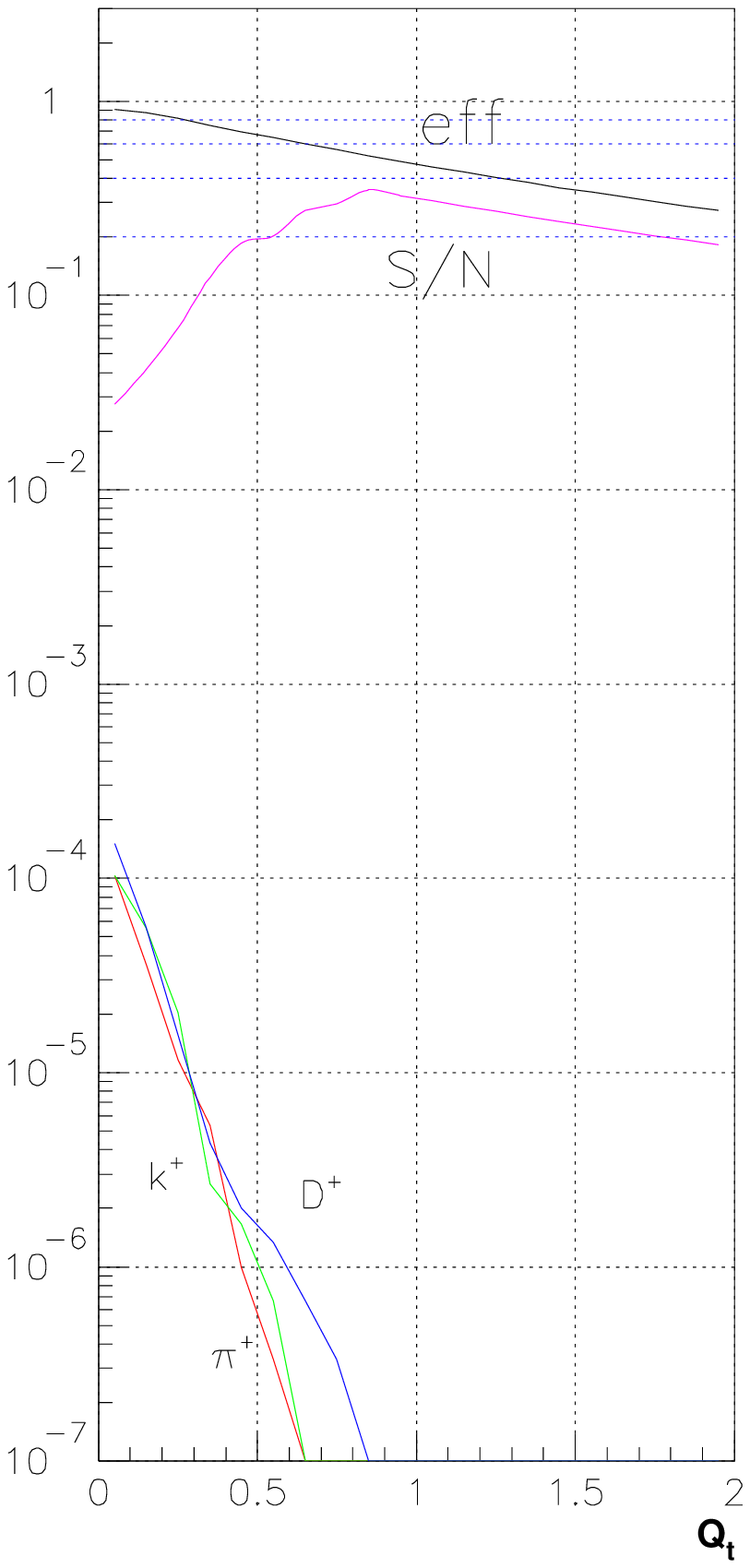,height=7cm,width=7cm}
}
\\
\numu Neutral Currents
\mbox{
\epsfig{file=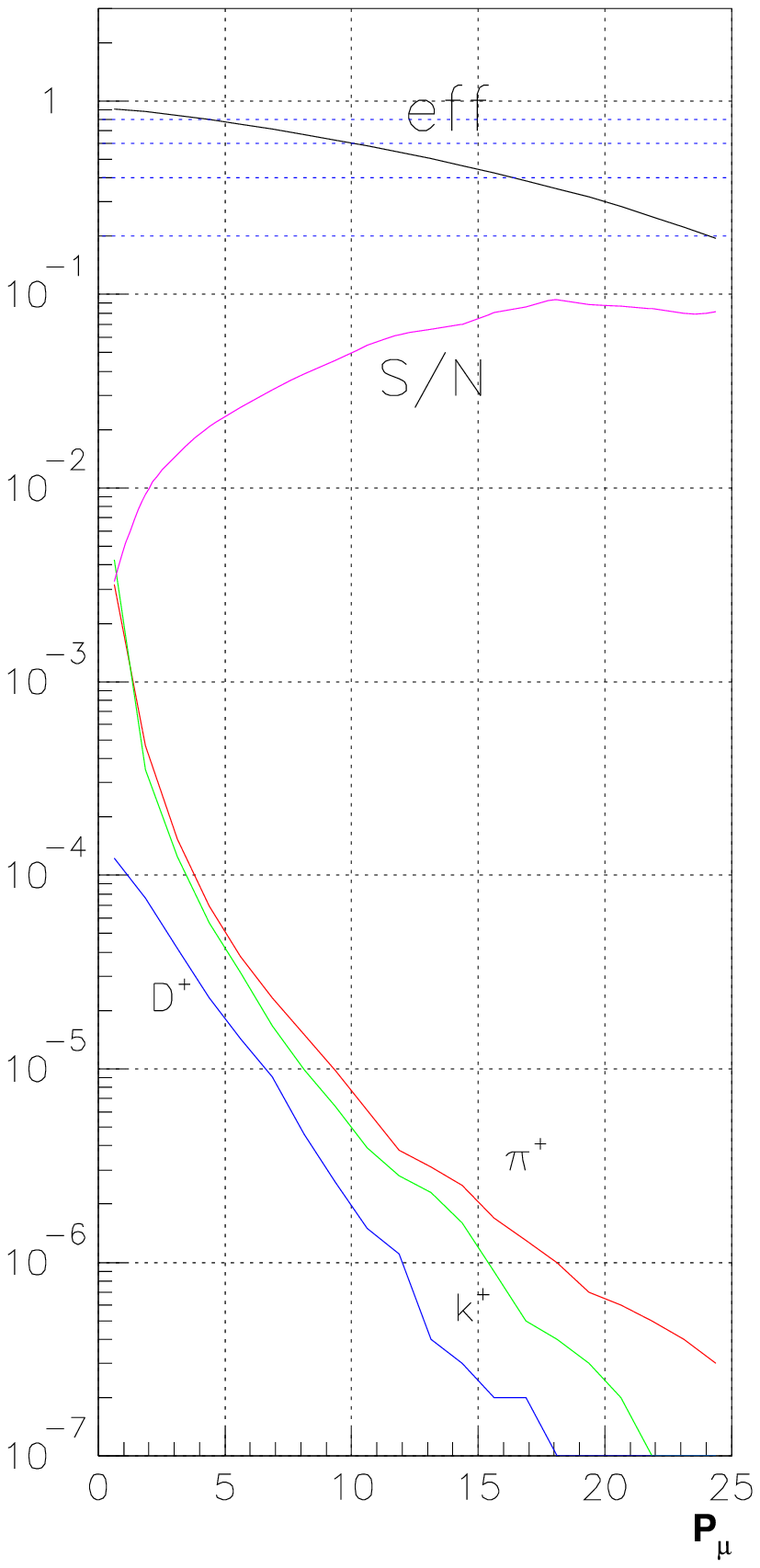,height=7cm,width=7cm} 
\epsfig{file=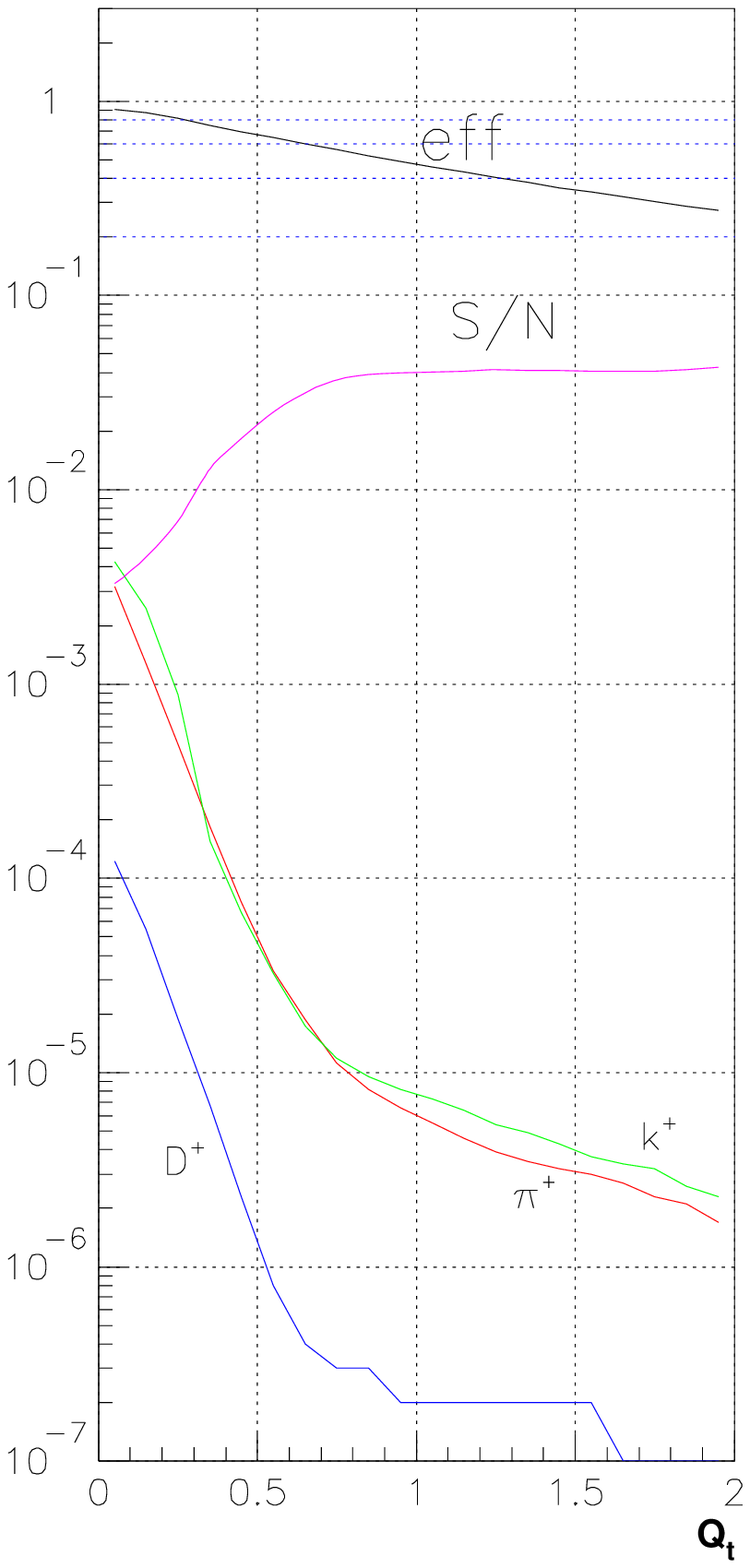,height=7cm,width=7cm}  
}    
\end{center}
\caption{\it Signal and backgrounds for \numu\ CC and NC events.}
\label{fig:ana2}
\end{figure}

The same exercise has to be repeated when a $\mu^-$ beam is considered.
The resulting neutrino beams  are now \numu\ and \nuebar\ and the
signal events are \numubar\ . 
Fig. \ref{fig:ana2} 
shows the efficiency for signal detection and
the fractional backgrounds, as a function of $P_\mu$ and $Q_t$, for
\numu\ charged and neutral currents. 
Tables \ref{tab:mums},
\ref{tab:mumeff} and \ref{tab:muminus}
summarize the results obtained (the cuts are the same
than for the $\mu^+$ analysis). Finally, Fig. \ref{fig:bins} shows the
signal efficiency and the fractional backgrounds for $\mu^+$'s and
$\mu^-$'s, as a function of the neutrino energy.

\begin{table}[bhtp]
\begin{center}
\begin{tabular}{|c|c|c|c|c|} \hline
  Baseline (km) & \numu\ CC   & \nuebar\ CC 
& \numu + \nuebar NC & 
\numubar\ (signal) 
\\ \hline \hline
732 & $6.9 \times 10^7$ & $3.0 \times 10^7$  & $3.1 \times 10^7$ & 
$5 \times 10^4$ \\
3500 & $3.0 \times 10^6$ & $1.3 \times 10^6$  & $1.4 \times 10^6$ & 
$1.6 \times 10^4$ \\
7332 & $6.9 \times 10^5$ & $3.0 \times 10^5$  & $3.1 \times 10^5$ & 
$0.2 \times 10^4$ \\
\hline
\end{tabular}
\end{center}
\caption{\it Data samples expected in a 40 kT detector for
$10^{21}$ useful $\mu^-$ decays.}
\label{tab:mums}
\end{table}
\begin{table}[bhtp]
\begin{center}
\begin{tabular}{|c|c|c|c|} \hline
    \numu\ CC   & \nuebar\ CC 
& \numu\ + \nuebar\ NC & 
\numubar\ (signal) 
\\ \hline \hline
$3.0 \times 10^{-7}$ & $6.0 \times 10^{-7}$  & $2.0 \times 10^{-6}$ & 
$4.0 \times 10^{-1}$ \\
\hline
\end{tabular}
\end{center}
\caption{\it Fractional backgrounds and signal selection efficiency for
the wrong-sign muon search with $\mu^-$ decays.}
\label{tab:mumeff}
\end{table}

\begin{table}[bhtp]
\begin{center}
\begin{tabular}{|c|c|c|c|c|} \hline
  Baseline (km) & \numu\ CC   & \nuebar\ CC 
& \numu\ + \nuebar\ NC & 
\numubar\ (signal) 
\\ \hline \hline
732 & $21 $ & $18$  & $66$ & 
$2.0 \times 10^4$ \\
3500 & $0.9$ & $0.8$  & $2.4$ & 
$6.4 \times 10^3$ \\
7332 & $0.2$ & $ 0.2 $  & $0.6$ & 
$8.0 \times 10^2$ \\
\hline
\end{tabular}
\end{center}
\caption{\it Events surviving the cuts in a 40 kT detector for
$10^{21}$ useful $\mu^-$ decays.}
\label{tab:muminus}
\end{table}

In summary, our study shows that a large magnetized iron calorimeter allows
the detection, with high efficiency ($\sim 30$--$40 \%$) of the 
golden-plated wrong-sign muon signal. The different backgrounds to this 
signal can be 
efficiently controlled using simple cuts, which exploit the different
kinematics between signal and background events. The charm background 
can be suppressed to {\it circa} $10^{-7}$, taking advantage of the high degree 
of collimation with the hadronic jet of charmed hadrons produced by 50 GeV 
muons. Instead, decays
of energetic pions or kaons in NC events contaminate the signal at 
the $10^{-6}$ level. The optimization of the $S/N$ ratio points to the   
intermediate distances of $O(3500$) km as the optimal baseline.


\section{Analysis in energy bins}
\label{bins}

A conservative estimate for the neutrino energy resolution in a detector
of the type described in the previous section 
is $\Delta E_\nu /E_\nu \sim 20\%$. 
For a $\mu$ beam of 50 GeV and the statistics of oscillated neutrinos expected 
in the range of parameters considered, it is reasonable to bin the 
data in five bins of 
equal width $\Delta E_\nu = 10$ GeV.

Let $N_{i,p}^\lambda$ be the total number of wrong-sign muons detected 
when the factory is run in 
polarity  $p=\mu^+,\mu^-$, grouped in 5 energy bins 
specified by $i = $ 1 to 5, and three possible 
distances, $\lambda =$  1 (732 km), 2 (3500 km), 3 (7332 km).

\begin{figure}
\begin{center}
\mbox{\epsfig{file=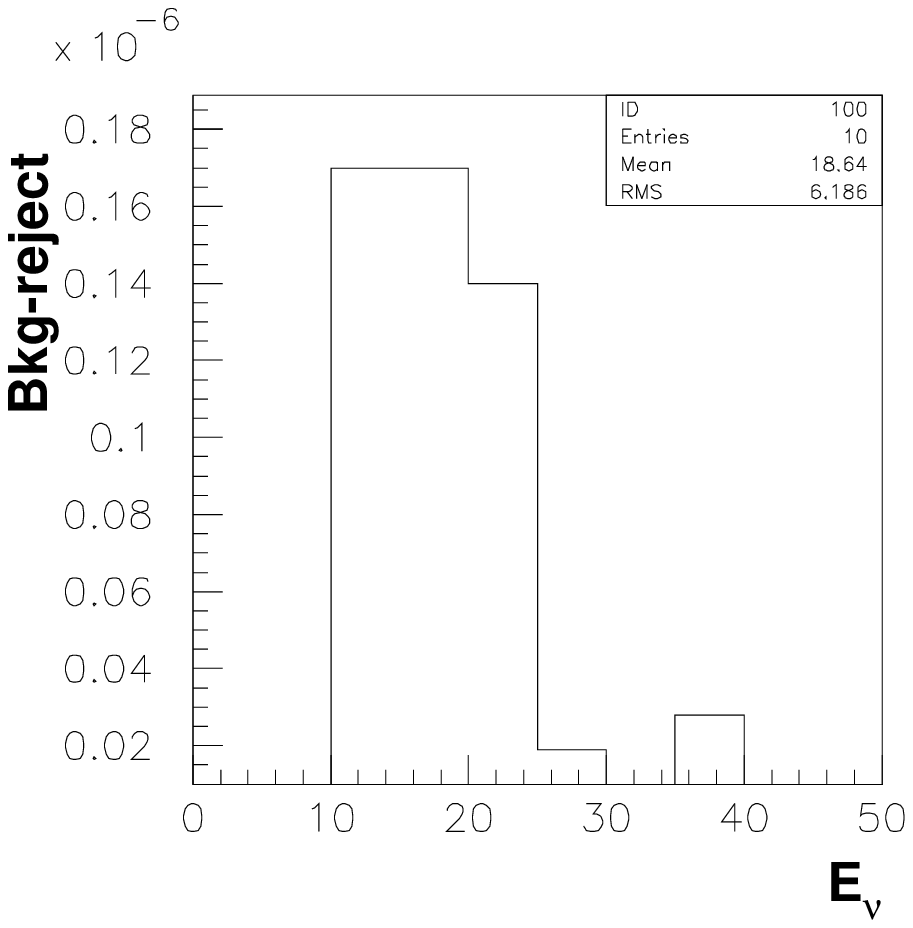,width=6cm} 
\epsfig{file=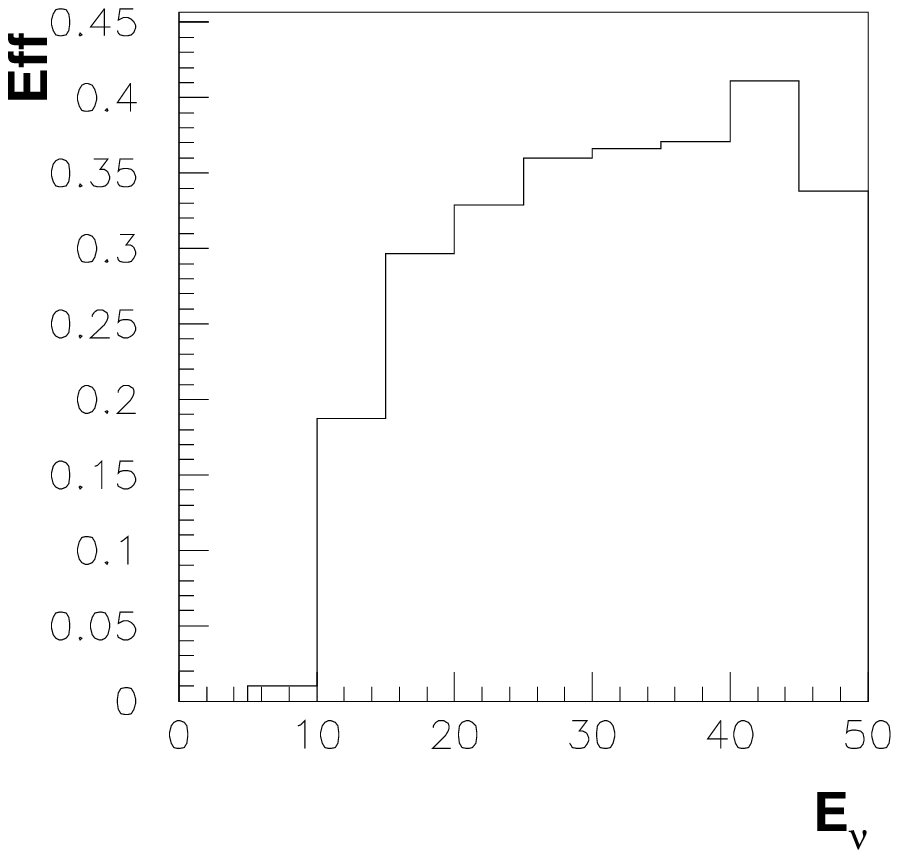,width=6cm}} \\ 
\mbox{\epsfig{file=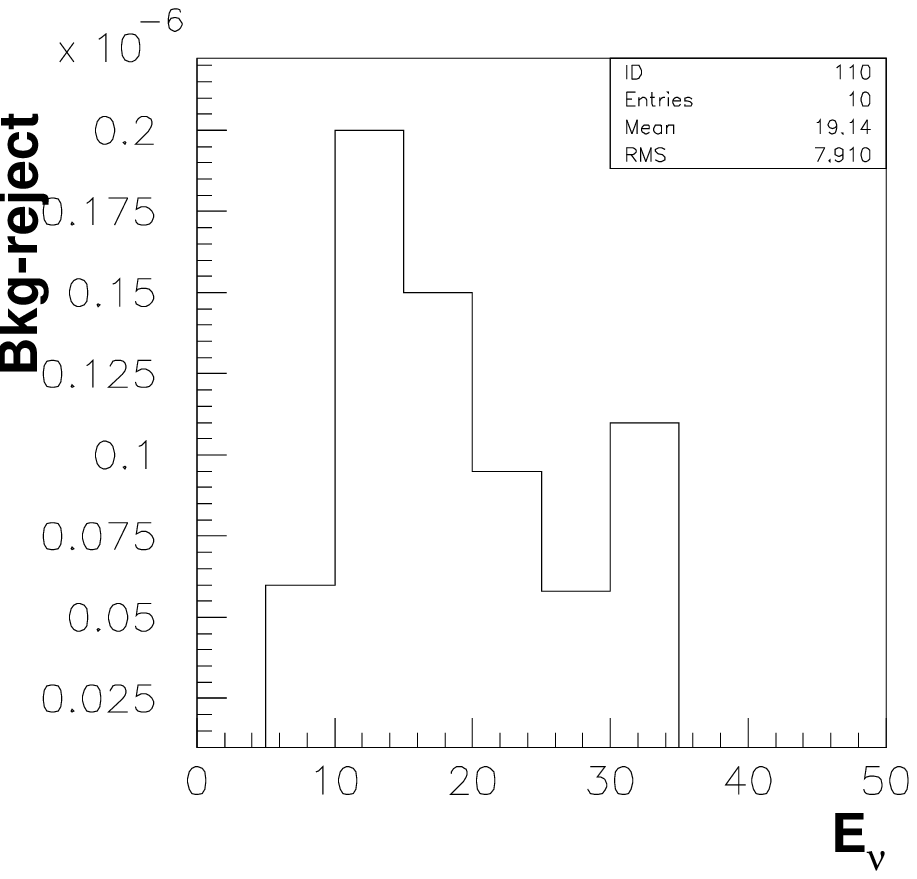,width=6cm} 
\epsfig{file=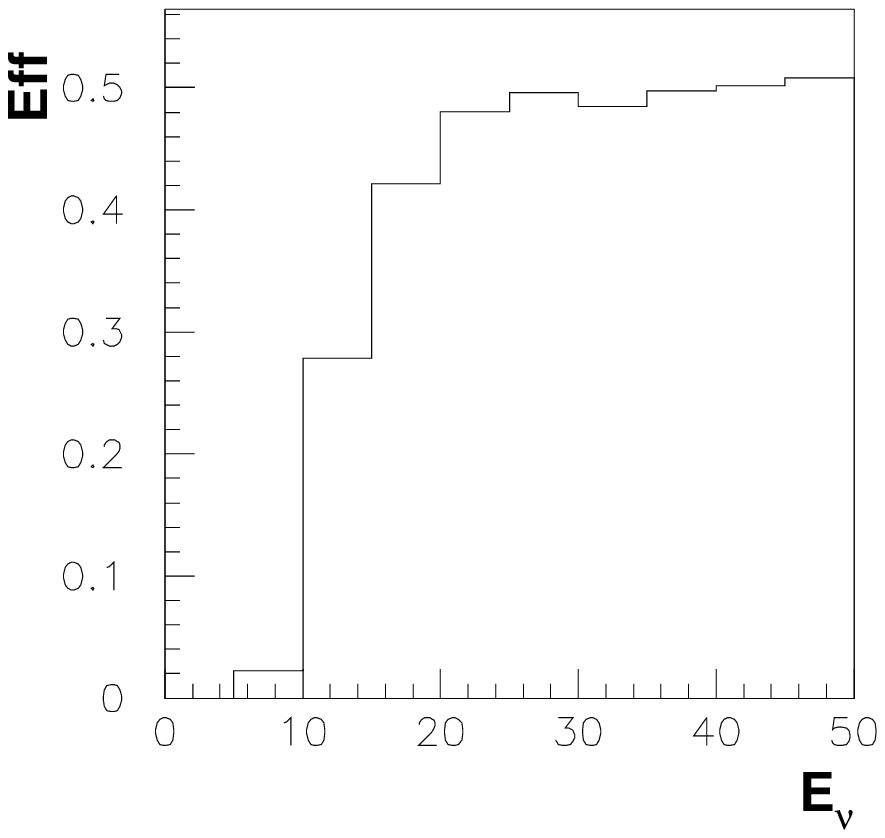,width=6cm}}
\end{center}
\caption{\it Signal efficiency and total fractional backgrounds for the polarities
$\mu^+$ (up) and $\mu^-$ (down) as a function of the neutrino energy.}
\label{fig:bins}
\end{figure}

In order to simulate a typical experimental situation we generate 
a set of ``data'' 
$n_{i,p}^\lambda$ as follows: for a given value of the oscillation parameters, 
the expected number of events, $N_{i,p}^\lambda$, is computed; taking into account backgrounds and 
detection efficiencies per bin, $b_{i,p}^\lambda$ and $\epsilon_{i,p}^\lambda$, 
as given in Fig.~\ref{fig:bins}, we then perform a Gaussian (or Poisson, depending on the number 
of events) smearing to mimic the statistical uncertainty. All in all, 
\begin{eqnarray} 
n_{i,p}^\lambda = \frac{ {\rm Smear} (N_{i,p}^\lambda \epsilon_{i,p}^\lambda + b_{i,p}^\lambda) - 
b_{i,p}^\lambda}{\epsilon_{i,p}^\lambda} \,. 
\end{eqnarray}

The ``data''  are then fitted to the theoretical expectation as a function 
of the neutrino parameters 
under study, using a $\chi^2$ minimization,
\be
\chi_\lambda^2 = \sum_p \sum_i 
\left(\frac{ n^\lambda_{i,p} \, - \, 
N^{\lambda}_{i,p}}{\delta n^\lambda_{i,p}}\right)^2 \, ,
\label{chi2}
\ee
where $\delta n^\lambda_{i,p}$ is the error of $n^\lambda_{i,p}$ (we include
no error in the efficiencies). 
We perform and  compare six different fits 
using: $\chi^2_1$, $\chi^2_2$, $\chi^2_3$ 
for the three distances, and the combinations $\chi^2_1 \, + \, \chi^2_2$, $\chi^2_2 \, + \,\chi^2_3$, 
$\chi^2_1 \, + \, \chi^2_2 \, + \,\chi^2_3$ to illustrate the gain in case the neutrino factory shoots 
to more than one location, a natural scenario given the ring 
configurations under study. 
For simplicity, we will consider a fit in at most two parameters at a time. 
All numerical results 
below will be obtained with the exact formulae for the oscillation probabilities.

\section{SMA-MSW or Vacuum solar deficit}
 
\begin{figure}
\begin{center}
\hskip -0.5cm
\mbox{\epsfig{file=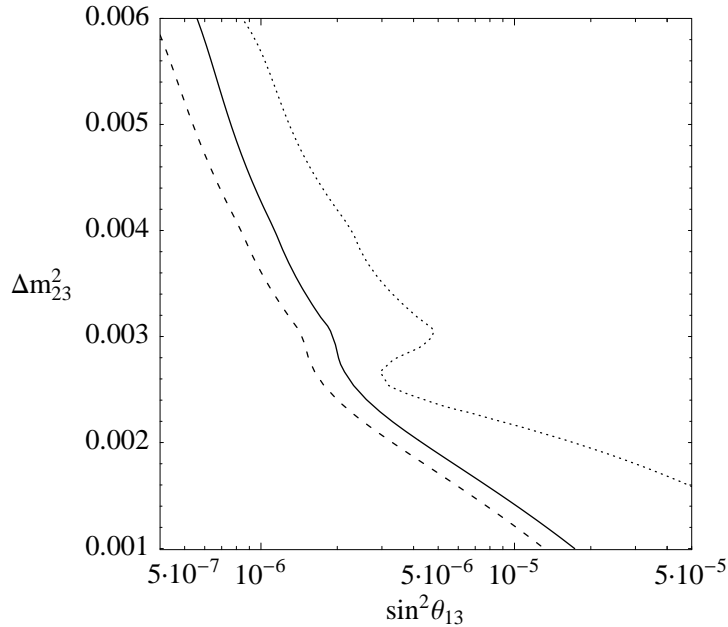,width=10cm} }
\end{center}
\caption{ \it Asymptotic sensitivity to $\sin^2 \tetaot$ as a function of 
$\Delta m^2_{23}$ at 90\% CL for $L=732$ km (dashed lines),
3500 km (solid lines) and 7332 km (dotted lines), in the SMA-MSW solution. 
Only stastistical errors are included.}  
\label{excluno}
\end{figure}

\begin{figure}
\begin{center}
\hskip -0.5cm
\mbox{\epsfig{file=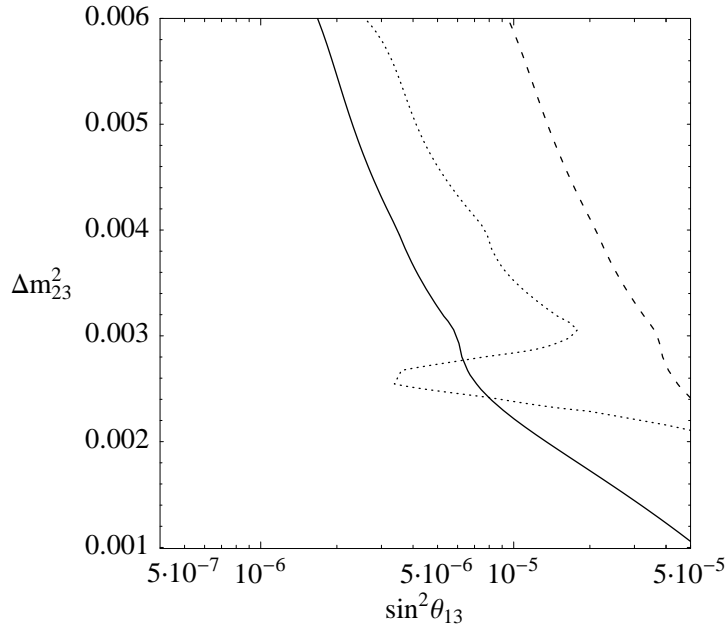,width=10cm} }
\end{center}
\caption{ \it As in Fig.~\ref{excluno}, including as well background
  errors and detection efficiencies.}
\label{exclusi}
\end{figure}

For the SMA-MSW or VO scenarios, the influence of solar parameters 
on the neutrino factory signals  
will be negligible\footnote{In practice, for the numerical results 
of this section, the central values 
in the SMA-MSW range are taken: $\Delta m_{12}^2 = 6 \times 10^{-6}$ eV$^2$ and 
$\sin^2 2 \theta_{12} = 0.006$.}, and CP-violation out of reach.
Besides its capability to reduce the errors on $\tetatt$ and $|\Delta m_{23}^2|$
to $\sim 1\%$ \cite{bgrw}, the factory would still be a unique machine to 
constrain/measure $\tetaot$ \cite{dgh} and the sign of $\Delta m^2_{23}$ 
\cite{Lipari:2000wy,ellis,bgrw}.

Consider first $\tetaot$. In Fig.~\ref{excluno}, we show the exclusion
plot 
at 90\% CL, on the  $\Delta m^2_{23}$ (in the range allowed by SuperK) versus $\sin^2 \tetaot$ plane, 
obtained with the full unbinned statistics and the two polarities. The
 same results, but including as well background errors and detection 
efficiencies are shown in Fig. \ref{exclusi}. The statistical
treatment 
in the presence of backgrounds is done as in 
\cite{feldmancousin}. Notice that the sensitivity is better at $L =$ 3500 km than at 732 km when 
efficiencies and backgrounds are included. The latter are responsible for it.
The sensitivity at $7332$ km is also worse than at $3500$ km, due to the loss in statistics.  
In conclusion, the sensitivity to $\tetaot$ can be improved by three-four orders of magnitude 
with respect to the present limits. This is consistent with the results of \cite{dgh} given the 
different statistics used.

The second major topic would be to perform the first precise measurements 
related to matter effects, in order to determine the sign 
of $\Delta m^2_{23}$ \cite{bgrw} 
and the size of the matter parameter, $A \propto n_e$.

We have studied the determination of the sign of $\Delta m^2_{23}$, assuming
that the absolute value has by then been measured with a precision of 10\%. 
We have explored the region around the best fit values of SuperKamiokande: 
$|\Delta m_{23}^2|=2.8 \times 10^{-3}$ eV$^2$ and $\tetatt=45^\circ$.
We perform a $\chi^2$ analysis on the $\Delta m^2_{23}, \tetaot$ plane, 
as described in last section. 
The conclusion is that, for ``data'' generated within the range 
$\tetaot = 1$--$10^\circ$ and 
$|\Delta m^2_{23}|$ in the range allowed by SuperKamiokande, a 
missidentification of the sign of $\Delta m^2_{23}$ can be
excluded at 99\% CL at 3500 km and 7332 km, but not at the shortest 
distance, 732 km. This conclusion agrees with the analysis of ref. \cite{bgrw}, 
which did not include the energy dependence information. 
 
\begin{figure}
\begin{center}
\mbox{\epsfig{file=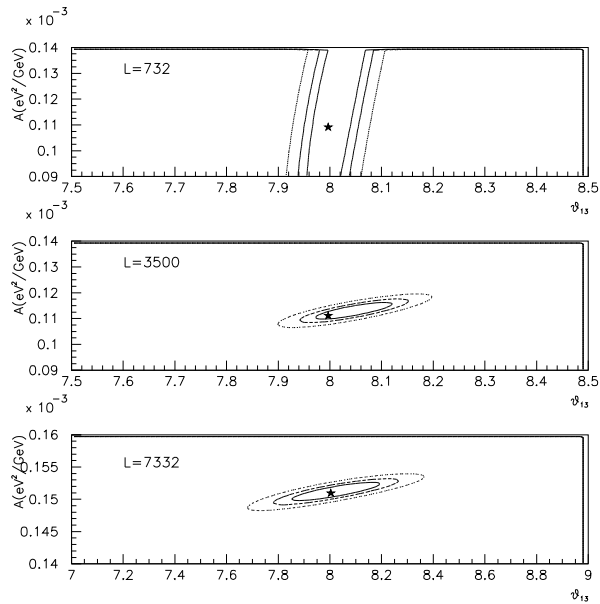,height=8cm}} 
\end{center}
\caption{\it 68.5, 90, 99 \% CL resulting from a simultaneous fit of 
$\tetaot$ and $A$. 
The parameters used to generate the ``data'' are denoted by a star, while
 the baseline(s) used in the 
fit is indicated in each plot. Only statistical errors included.}
\label{fig:sma8}
\end{figure}

\begin{figure}
\begin{center}
\mbox{\epsfig{file=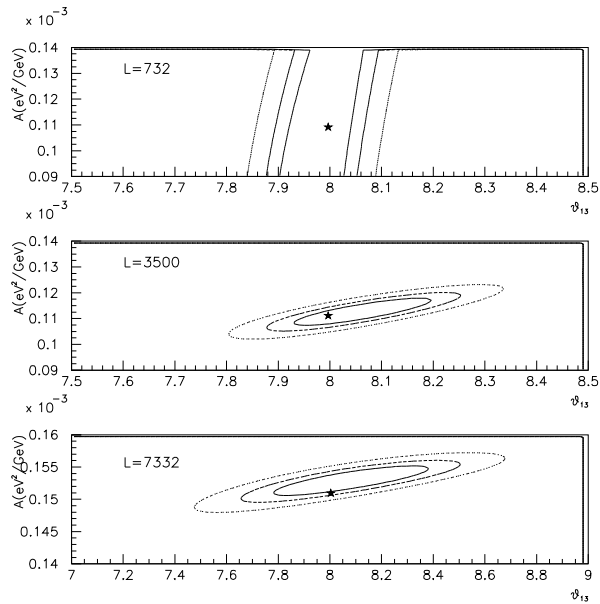,height=8cm}} 
\end{center}
\caption{\it The same as in Fig.~\ref{fig:sma8} but including backgrounds and efficiencies.}
\label{fig:sma8be}
\end{figure}

We have further studied how the matter parameter of Eq.~(\ref{a}) and the angle $\tetaot$ can be measured simultaneously.  
Fig.~\ref{fig:sma8} shows the result of a $\chi^2$ fit as described 
in section \ref{bins}.  
Only statistical errors have been included in this figure.
The corresponding results including as well background errors and detection efficiencies 
are shown in Fig.~\ref{fig:sma8be}. At 732 km there is no sensitivity to the matter term, as expected. 
However, already at 3500 km, $A$ can be measured with a 10\% precision. 
At the largest baseline, 
the precision in $A$ improves although at the expense of loosing precision 
in $\tetaot$ due to the loss in statistics. The level of precision discussed 
here might even be interesting for geophysicists \cite{geo}.

The above conclusions also hold for the vacuum solution to the solar deficit.

\section{LMA-MSW}
\label{lma}

Assume now the LMA-MSW scenario. 
Fixed values of the atmospheric parameters are used in this section, 
$\Delta m_{23}^2 = 2.8\times 10^{-3}$ eV$^2$ and maximal mixing,
$\tetatt=45^\circ$. A precision of $1\%$ in these parameters is
achievable through  
  muon 
disappearance measurements at the neutrino factory \cite{bgrw}. This level
of uncertainty is not expected to affect the results of this section. 

Let us start discussing the measurement  of the CP phase $\delta$ versus $\tetaot$.
We have studied numerically how to disentangle them
in the range $1$--$10^\circ$ 
and $0 \le \delta \le 180^\circ$.

\begin{figure}
\begin{center}
\epsfig{file=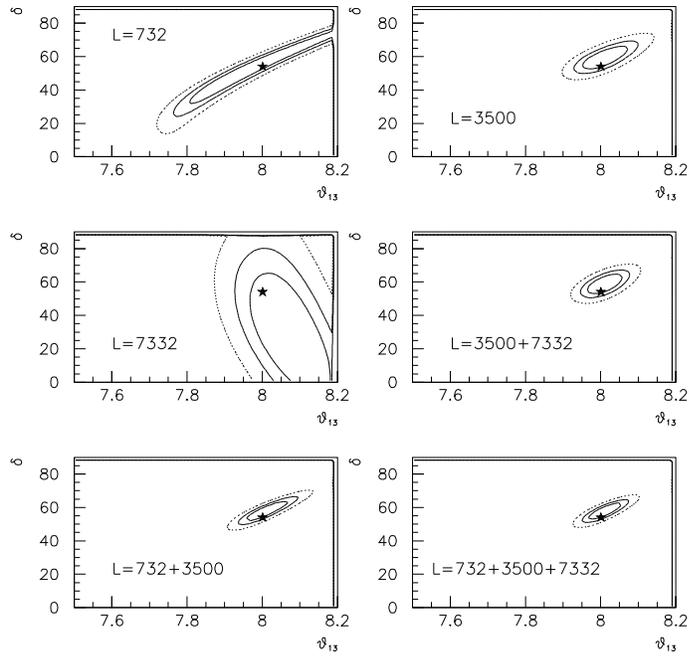,width=10cm} 
\end{center}
\caption{\it 68.5, 90, 99 \% CL contours resulting from a $\chi^2$ fit of  $\tetaot$ and $\delta$. 
The parameters used to generate the ``data'' are depicted by a star and the baseline(s) which is used for 
the fit indicated in each plot. Only statistical errors are included.}
\label{fig:54}
\end{figure}

\begin{figure}
\begin{center}
\epsfig{file=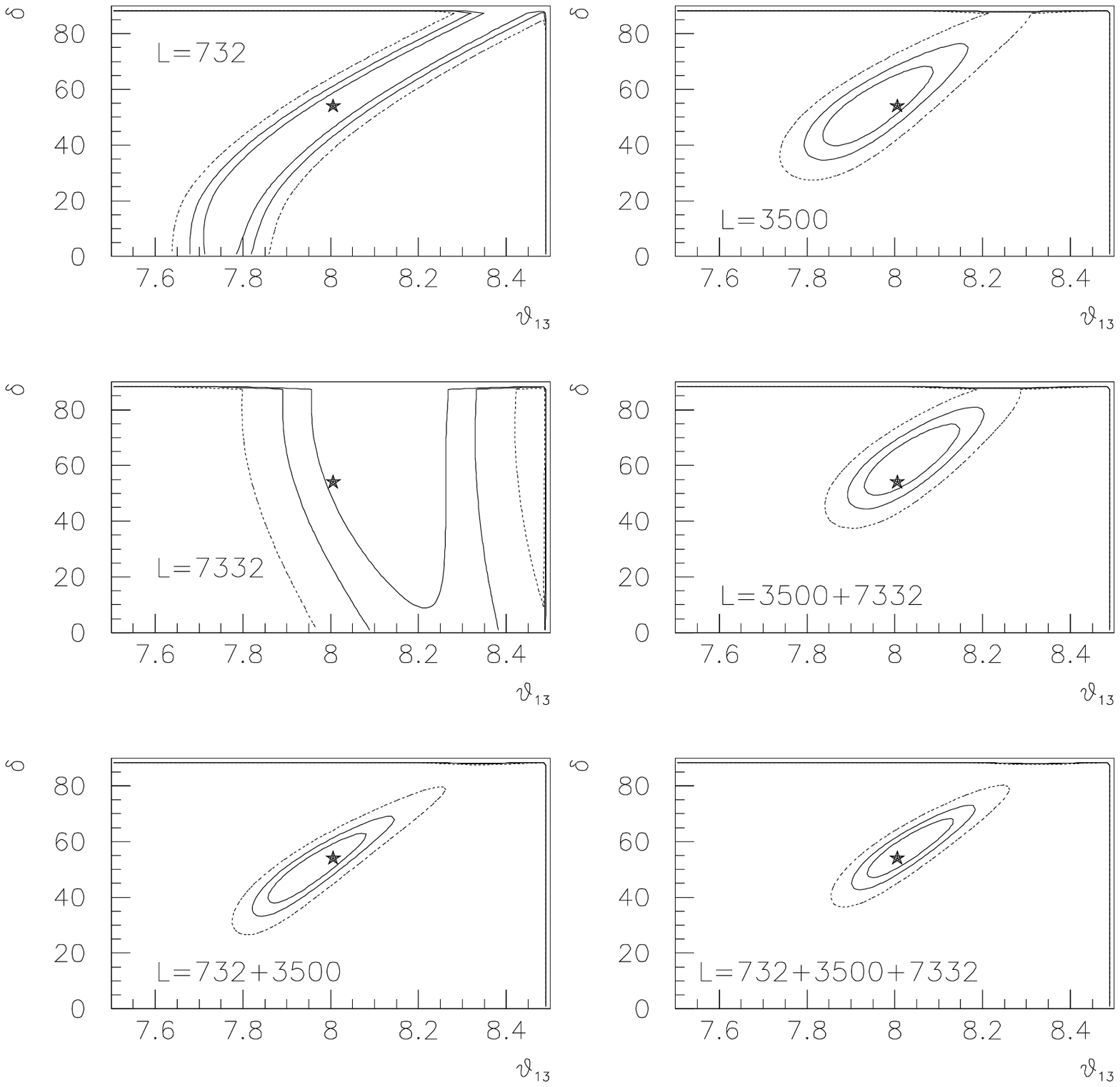,width=10cm} 
\end{center}
\caption{\it The same as Fig.~\ref{fig:54} with backgrounds and efficiencies included.}
\label{fig:demonios}
\end{figure}

Consider first the upper solar mass range 
allowed by the LMA-MSW solution: 
$\Delta m_{12}^2=10^{-4}$ eV$^2$. Fig.~\ref{fig:54} shows the confidence 
level contours for a simultaneous fit
of $\tetaot$ and $\delta$, for ``data'' corresponding to
 $\tetaot=8^\circ$, $\delta=54^\circ$, including only 
statistical errors in the analysis. 
Fig.~\ref{fig:demonios} shows the same analysis 
taking into account the 
background errors and detection efficiencies of Fig.~\ref{fig:bins}. 
The correlation between $\delta$ and 
$\tetaot$ is very large at the shortest baseline 732 km, as argued 
in section \ref{sec:prob}. 
The phase $\delta$ is then not measurable and this indetermination 
induces a rather large error on the 
angle $\tetaot$. However, at the intermediate baseline of 3500 km the two 
parameters can be disentangled 
and measured. At the largest baseline, the sensitivity to $\delta$ is lost 
and the precision in 
$\tetaot$ becomes worse due to the smaller statistics. The combination of 
the results for 3500 km 
with that for any one of the other distances improves the fit, but not in 
a dramatic way.
Just one detector placed at ${\cal O}$ (3000 km) may be sufficient: a 
precision of few tenths of degree
is attained for $\tetaot$ and of a few tens of degrees for $\delta$.

\begin{figure}
\begin{center}
\epsfig{file=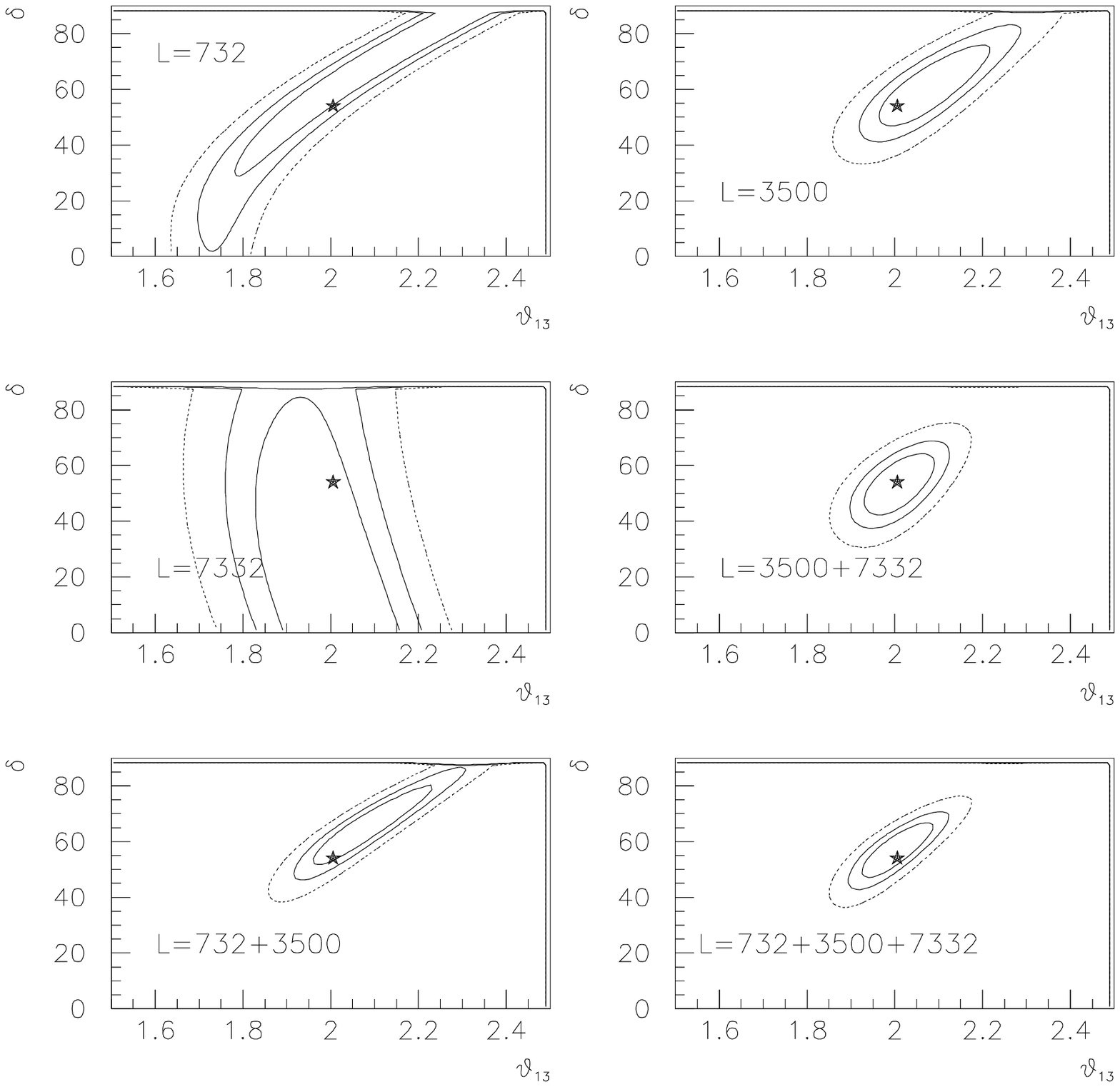,width=10cm} 
\end{center}
\caption{\it As Fig.~(\ref{fig:demonios}) for different values of the
parameters denoted by the star. }
\label{fig:54_2_be}
\end{figure}

\begin{figure}
\begin{center}
\epsfig{file=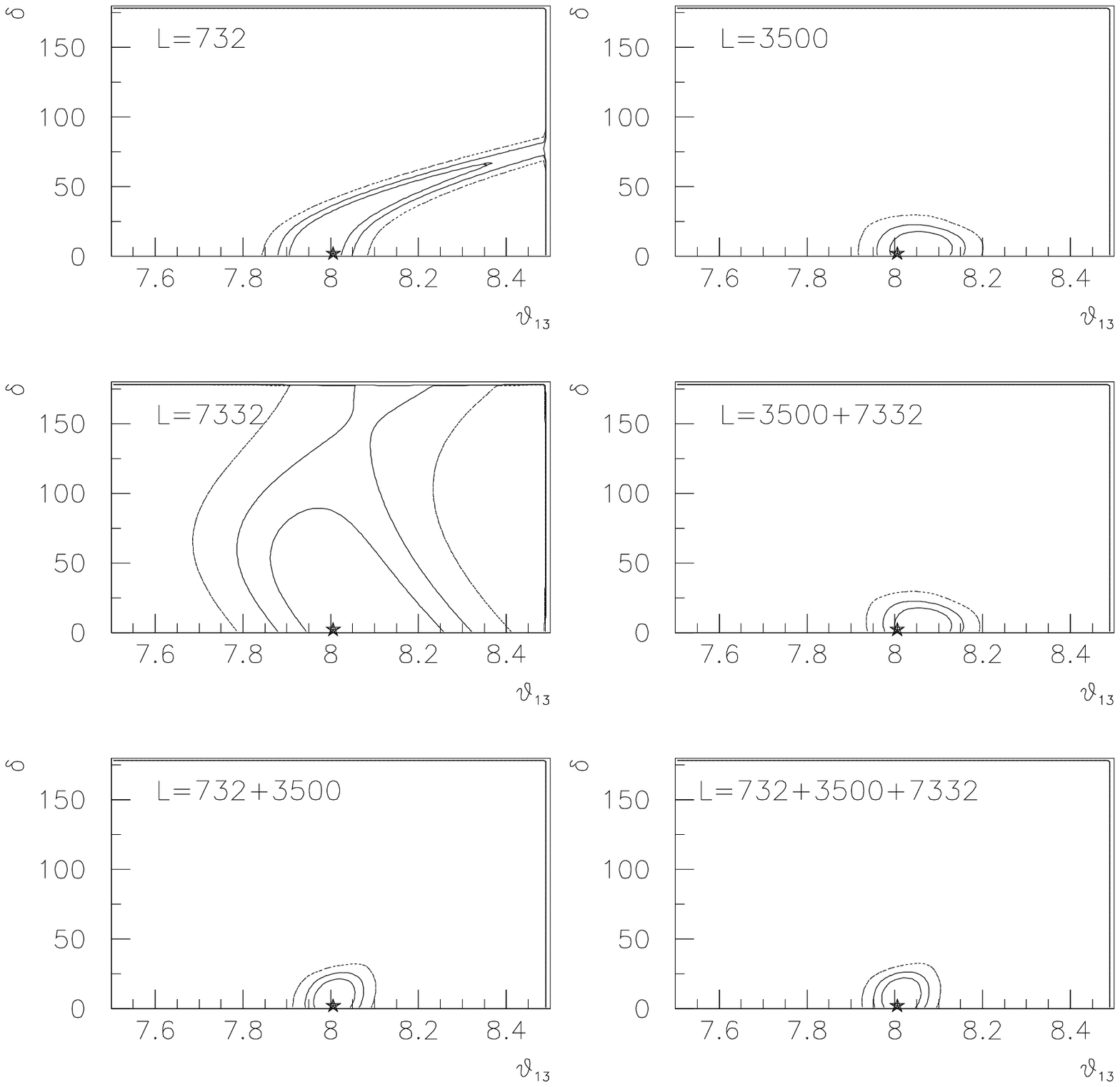,width=10cm} 
\end{center}
\caption{\it As Fig.~(\ref{fig:demonios}) for different values of the
parameters denoted by the star.}
\label{fig:00_8_be}
\end{figure}

Similar figures are obtained for ``data'' corresponding to 
 smaller values of $\tetaot$, as 
shown in Fig.~\ref{fig:54_2_be} for  
$\tetaot=2^\circ$. The pattern is maintained as well for 
different values of $\delta$: see  
Fig.~\ref{fig:00_8_be} for ``data'' corresponding to 
$\delta=0^\circ$ and $\tetaot=8^\circ$. 
This last figure also proves 
that, if the sign of $\Delta m_{12}^2$ is known by the time the 
neutrino factory will be operative,
$\delta=0^\circ$ is distinguishable from $\delta=180^\circ$ with 
just one baseline. 
This exemplifies the power of the analysis of the energy dependence. 
Recall in any case that a $\pi$-ambiguity on $\delta$ has no bearing 
on the existence of CP-violation, and we will go on 
considering positive values of $\Delta m_{12}^2$.


\begin{figure}
\begin{center}
\epsfig{file=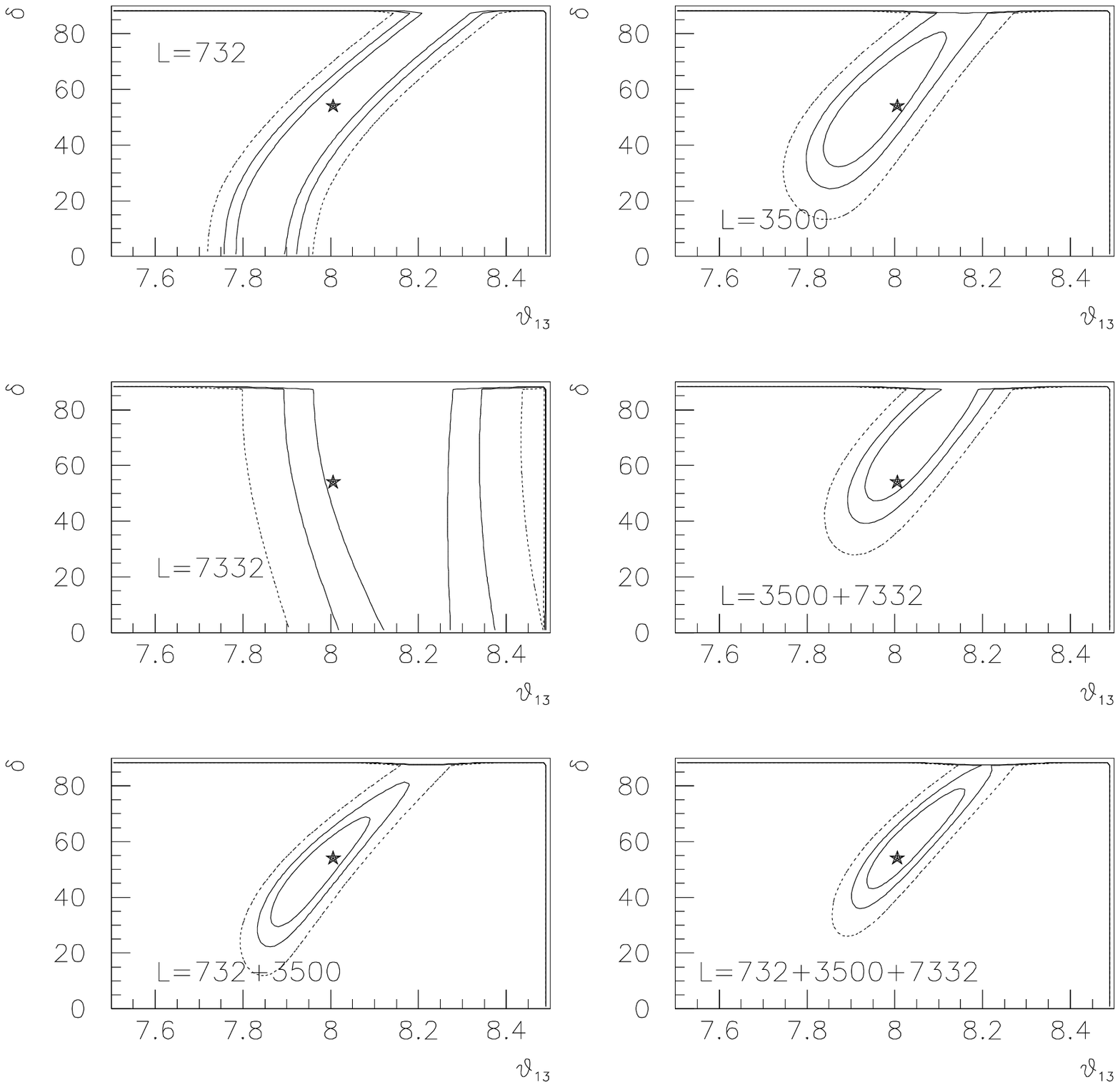,width=10cm} 
\end{center}
\caption{ \it Fit for $\Delta m_{12}^2 = 5 \times 10^{-5}$ eV$^2$ including backgrounds errors and 
detection efficiencies. The star indicates the parameters used to generate
the ``data''.}
\label{fig:54_8_be_bb}
\end{figure}

\begin{figure}
\begin{center}
\epsfig{file=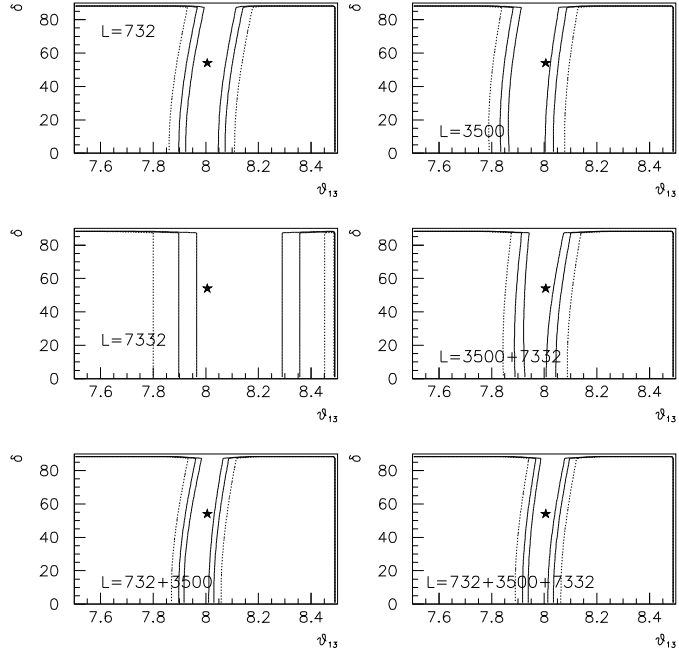,width=10cm} 
\end{center}
\caption{\it As Fig.~\ref{fig:54_8_be_bb}, for $\Delta m_{12}^2 = 1 \times 10^{-5}$ eV$^2$.}
\label{fig:54_8_be_b}
\end{figure}

The sensitivity to CP-violation decreases linearly
with $\Delta m_{12}^2$. At the central value allowed by the LMA-MSW solution, 
$\Delta m_{12}^2=5\times 10^{-5}$ eV$^2$,  CP-violation can still 
be discovered, as shown 
in Fig.~\ref{fig:54_8_be_bb}. At the lower value allowed, 
$\Delta m_{12}^2 = 1 \times 10^{-5}$ eV$^2$, 
the sensitivity to CP-violation is lost with the experimental set-up considered, 
as shown in Fig.~\ref{fig:54_8_be_b}.

\begin{figure}
\begin{center}
\epsfig{file=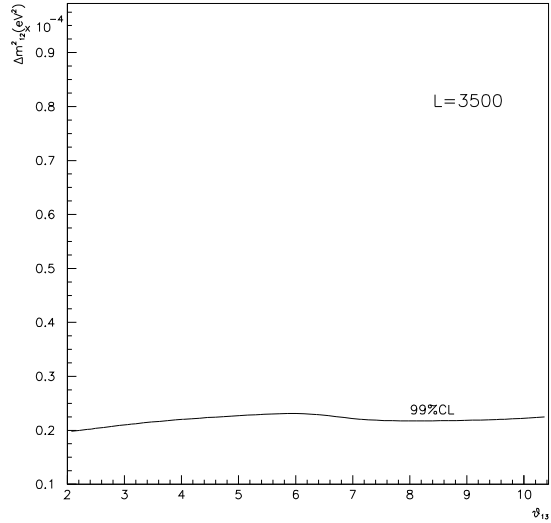,width=8cm}
\end{center} 
\caption{\it Lower limit in $\Delta m_{12}^2$ at which a maximal CP phase ($90^\circ$) can be distinguished
from a vanishing phase at 99\% CL, as a function of $\tetaot$ at $L = $ 3500
km. Background errors and 
efficiencies are included.}
\label{fig:limdm12}
\end{figure}

We have quantified what is the minimum value of $\Delta m^2_{12}$ for which 
a maximal CP-odd phase, $\delta = 90^\circ$, can be distinguished at 99\% CL from $\delta = 0^\circ$. 
The result is shown in Fig.~\ref{fig:limdm12}: $\Delta m^2_{12} > 2 \times 10^{-5}$ eV$^2$, 
with very small dependence on $\tetaot$, in the range considered. 

\begin{figure}
\begin{center}
\epsfig{file=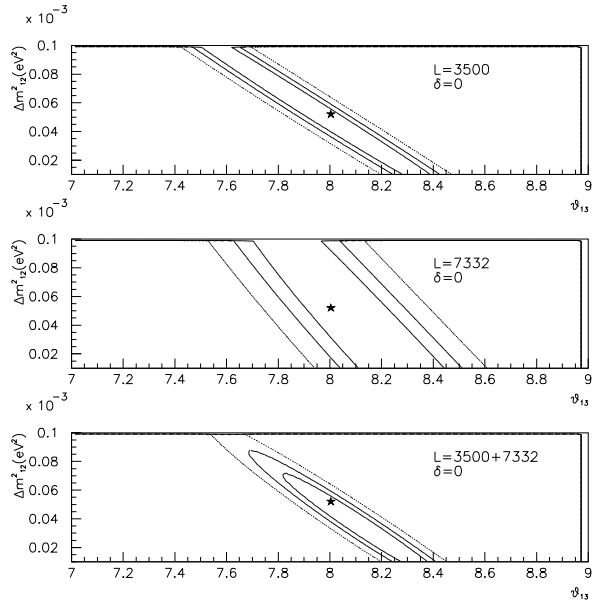,height=8cm} 
\end{center}
\caption{\it Simultaneous fit to $\tetaot$ and $\Delta m_{12}^2$. 
The range shown in the vertical axis 
is the presently allowed LMA-MSW range. The star indicates the parameters used
to generate the ``data'' and the CP-odd phase is set to zero. 
Backgrounds  and detection efficiencies are included.}
\label{fig:mmbe}
\end{figure}

One word of caution is pertinent: up to now we assumed $|\Delta m^2_{12}|$ 
and $\sin 2 \theta_{12}$ known by the time the neutrino factory will be 
operational. 
Otherwise, the correlation of these parameters with $\tetaot$ would be 
even more problematic than that between 
$\delta$ and $\tetaot$, as illustrated in Fig.~\ref{fig:mmbe} 
for $|\Delta m^2_{12}|$. 
The error induced on the measurement of $\tetaot$ by the present uncertainty  
in $|\Delta m^2_{12}|$ is much larger than that stemming from the 
uncertainty on $\delta$. 
Fortunately, LBL reactor experiments will measure $|\Delta m^2_{12}|$ 
and $\sin 2 \theta_{12}$ if it lies  in the LMA-MSW range. 
Even if the error in these measurements is as large as 50\%, 
the problem would be much less serious. 
We have checked that such uncertainty  does
not affect our results concerning the sensitivity to $\delta$, and only
induces an error in $\theta_{13}$ that can be read from Fig.~\ref{fig:mmbe}.  

Concerning the measurement of the matter parameter, we have considered 
simultaneous fits of $\tetaot$ and $A$, for two values of the
CP-phase: $\delta=0, \pi/2$.
 The confidence level contours obtained are very similar to those 
in Fig. \ref{fig:sma8} for the SMA-MSW solution. This indicates that there 
is no dangerous correlation between $A$ and $\tetaot$ in the presence of
sizeable $\delta$-dependent terms, and both parameters can be 
safely measured at 3500 km. However, it is 
important to stress that the simultaneous measurement of the three parameters 
$\theta_{13}, \delta$  and $A$ will increase the errors with respect to 
the two-parameter fits performed here. In this respect the combination 
of two baselines: ${\cal O}$(3500 km) and ${\cal O}$(7332 km) may be helpful. 

\newpage

\section{Summary}

The neutrino beams obtained from muon storage rings will be excellent for 
precision neutrino physics.
The appearance of wrong-sign muons is a powerful neutrino oscillation signal, 
which allows to improve considerably our knowledge of the leptonic flavour
sector.

Two very important questions are the optimal energy for the decaying 
muons and the optimal detection 
distance(s), in view of the physics goals. The higher the parent 
muon energy, the larger the oscillation signals. This fact, together with the 
requirement of low backgrounds 
and good detection efficiencies, lead to consider muon energies as 
high as possible within realistic 
machine designs. Energies of a several tens of GeV are currently 
under discussion, assumed here to 
be $E_\mu =$ 50 GeV, for definiteness.

Energy and detection distance are intertwined in the oscillation pattern of
neutrinos propagating in  matter.
We have derived an analytical approximate formula for the oscillation 
probabilities in matter, which helps to understand how the sensitivity to
the most interesting quantities scales with the neutrino energy and distance.

\begin{figure}
\begin{center}
\epsfig{file=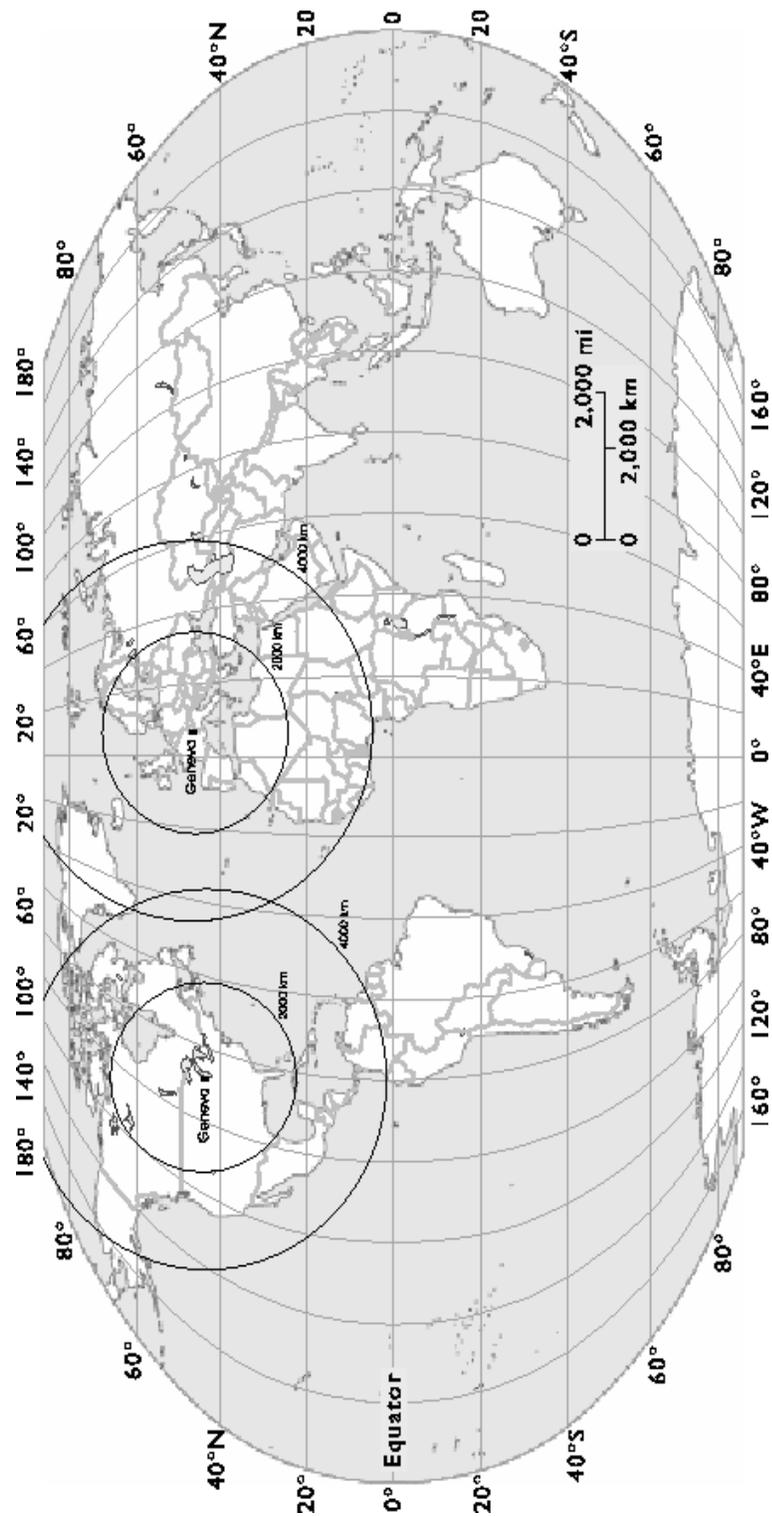,angle=90,height=20cm,width=10cm} 
\end{center}
\caption{\it 
Artistic view of possible baselines of ${\cal O}$(3000 km) at which a
neutrino factory, located at Geneva (Switzerland) or Geneva (Illinois),
could shoot. The inner and outer circles correspond to 
2000 and 4000 km, respectively. Note that the circles have not 
been projected onto the Earth surface.}
\label{fig:world}
\end{figure}

We have shown that an analysis in neutrino energy bins, combined 
with a comparison of the signals obtained with the two polarities, 
allows to disentangle the unknown parameters at long enough baselines.
In particular, for the LMA-MSW solution, $\tetaot$ and $\delta$
can be simultaneously measured. We have also studied realistic 
backgrounds and detection efficiencies. The overall conclusion is 
that the intermediate baseline of ${\cal O}$(3000 km) is optimal for the 
physics goals considered in this paper (see Fig.~\ref{fig:world} for 
an artistic view of possible locations). 

Quantitatively, our two parameter fits at 3500 km indicate:

\begin{itemize}
\item 
The angle $\tetaot$ can be measured with a precision of tenths of degrees, 
down to values of $\tetaot=1^\circ$. The asymptotic sensitivity to 
$\sin^2 \tetaot$ can be improved by three orders of magnitude or more.

\item 
If the solar deficit corresponds to solar parameters in the LMA-MSW range, 
CP-violation may be tackled. The phase $\delta$ can be determined with a 
precision of tens of degrees, for the central values allowed 
for $|\Delta m_{12}^2|$, and  maximal CP-violation 
can be disentangled from no CP-violation  at 99\% CL
for values of $|\Delta m_{12}^2| > 2 \times 10^{-5}$ eV$^2$.

\item The sign of the atmospheric mass difference, $\Delta m_{23}^2$, can be 
determined at 99\% CL, for $\tetaot$ within the 
range $\tetaot=1$--$10^\circ$ and $|\Delta m_{23}^2|$ 
in the range allowed by SuperKamiokande data. 

\item A model independent confirmation of the MSW effect will be feasible, 
and the matter parameter $A$ measured within a 10\% precision, or better
if combined with the longest baseline: $7332$ km. 

\end{itemize}

In the case of the LMA-MSW solution, the combination of the two
longest baselines may be useful if a multiparameter fit becomes necessary.

Even though non-zero neutrino masses are barely established, the
neutrino 
sector of the theory can be 
convincingly argued to herald physics well beyond the standard model.
It is in this perspective that a neutrino factory should be built.

%
\section{Acknowledgements}
We acknowledge useful conversations with:
J.~Bernabeu, A.~de~Gouvea,
A.~De~R\'ujula, F.~Dydak, J.~Ellis and C.~Quigg. 
We further thank A.~De~R\'ujula for suggestions and a critical reading of the
manuscript. We thank P.~Lipari for pointing out an incorrect statement 
in an earlier version of this paper. 
A.~D. acknowledges the I.N.F.N. for financial support. S.~R. 
acknowledges the European Union for financial support 
through contract ERBFMBICT972474. The work of A.~D., M.~B.~G. and S.~R. 
was also partially supported by CICYT project AEN/97/1678. 

\newpage

{\Large{\bf Appendix A: Non-oscillated statistics}} 

Tab.~\ref{tab:tab2} contains the results for the $\nu_e$ and $\bar \nu_\mu$ fluxes for the parent 
muon energy $E_\mu = $ 50 GeV and for $n_\mu = 2 \times 10^{20}$ useful muons per
year, per 5 operational years. The result for three muon polarizations are shown: 
${\cal P}_{\mu^\pm} =$ 0, $\mp$ 0.3 (the ``natural'' polarization) and $\mp 1$. 
For ${\cal P}_{\mu^\pm} =0$ our results agree with \cite{bgw} when the same number of useful muons 
is considered.

For a $\mu^+$ ($\mu^-$) beam of $E_\mu=50$ GeV, the average neutrino and antineutrino energies are,
for ${\cal P}_{\mu^+(\mu^-)} = 0$, $\langle E_{\bar \nu_\mu ( \nu_\mu)} \rangle =$ 35 GeV,
and $\langle E_{ \nu_e (\bar \nu_e)} \rangle =$ 30 GeV.
For ${\cal P}_{\mu^+(\mu^-)} = -1 ( +1)$, we get $\langle E_{\bar \nu_\mu(\nu_\mu)}\rangle =$ 30 GeV,
$\langle E_{ \nu_e (\bar \nu_e)} \rangle =$ 30 GeV.  

\begin{table}[bht]
\vskip 1cm
\centering
\begin{tabular}{||c|c||c|c||c|c||}
\hline \hline
\multicolumn{2}{||c||}{$E_{\mu^{\mp}}= 50$ GeV} & 
\multicolumn{2}{c||} {$\mu^{-}$} &  
\multicolumn{2}{c||}{$\mu^{+}$}\\
\hline
\multicolumn{2}{||c||}{L (km)} & $\Phi_{\nu_\mu}/10^{12}$ & 
$\Phi_{\bar\nu_e}/10^{12}$  & $\Phi_{\bar\nu_\mu}/10^{12}$ & 
$\Phi_{\nu_e}/10^{12}$ \\  
\hline\hline
\multicolumn{2}{||c||}{732}& \multicolumn{2}{r} {} & 
\multicolumn{2}{l||} {} \\
\hline\hline
   & $0$       & $123$ & $122$ & $123$ & $122$\\
\cline{2-6}
$\cal{P}_{\mu^{\mp}}$
   & ${\pm}0.3$& $110$ & $159$ & $110$ & $159$\\
\cline{2-6}
   & ${\pm}1$  & $81.5$ & $244$ & $81.5$ & $244$\\ 
\hline\hline
\multicolumn{2}{||c||}{3500}& \multicolumn{2}{r} {} & 
\multicolumn{2}{l||} {} \\
\hline\hline
   & $0$       & $5.35$ & $5.35$ & $5.35$ & $5.35$\\
\cline{2-6}
$\cal{P}_{\mu^{\mp}}$
   & ${\pm}0.3$& $4.83$ & $6.95$ & $4.83$ & $6.95$\\
\cline{2-6}
   & ${\pm}1$  & $3.56$ & $10.7$ & $3.56$ & $10.7$\\ 
\hline\hline
\multicolumn{2}{||c||}{7332}& \multicolumn{2}{r} {} & 
\multicolumn{2}{l||} {} \\
\hline\hline
   & $0$       & $1.22$ & $1.22$ & $1.22$ & $1.22$\\
\cline{2-6}
$\cal{P}_{\mu^{\mp}}$
   & ${\pm}0.3$& $1.10$ & $1.58$ & $1.10$ & $1.58$\\
\cline{2-6}
   & ${\pm}1$  & $0.81$ & $2.43$ & $0.81$ & $2.43$\\ 
\hline\hline
\end{tabular}
\caption{\it 
Neutrino and antineutrino fluxes for L = 732, 3500 and 7332 Km per m$^2$ 
per 5 operational years when $2 \times 10^{20}$  muons decay in the straight 
section of the storage ring. This fluxes have been averaged over an angular 
divergence of 0.1 mr.}
\label{tab:tab2}
\end{table}

\begin{table}[thb]
\vskip 1cm
\centering
\begin{tabular}{||c|c||c|c||c|c||}
\hline \hline
\multicolumn{2}{||c||}{$E_{\mu^{\mp}}= 50$ GeV} & 
\multicolumn{2}{c||} {$\mu^{-}$} &  
\multicolumn{2}{c||}{$\mu^{+}$}\\
\hline
\multicolumn{2}{||c||}{L (km)} & $N_{\nu_\mu}/10^5$ & $N_{\bar\nu_e}/10^5$  & 
$N_{\bar\nu_\mu}/10^5$ &$N_{\nu_e}/10^5$ \\  
\hline\hline
\multicolumn{2}{||c||}{732}& \multicolumn{2}{r} {} & 
\multicolumn{2}{l||} {} \\
\hline\hline
   & $0$       & $692$ & $300$ & $352$ & $590$\\
\cline{2-6}
$\cal{P}_{\mu^{\mp}}$
   & ${\pm}0.3$& $603$ & $390$ & $306$ & $768$\\
\cline{2-6}
   & ${\pm}1$  & $394$ & $600$ & $200$ & $1180$\\ 
\hline\hline
\multicolumn{2}{||c||}{3500} & \multicolumn{2}{r} {} & 
\multicolumn{2}{l||} {} \\
\hline\hline
   & $0$       & $30.4$ & $13.1$ & $15.4$ & $25.8$\\
\cline{2-6}
$\cal{P}_{\mu^{\mp}}$
   & ${\pm}0.3$& $26.4$ & $17.1$ & $13.4$ & $33.6$\\
\cline{2-6}
   & ${\pm}1$  & $17.2$ & $26.2$ & $8.75$ & $51.6$\\ 
\hline\hline
\multicolumn{2}{||c||}{7332}& \multicolumn{2}{r} {} & 
\multicolumn{2}{l||} {} \\
\hline\hline
   & $0$       & $6.90$ & $3.00$ & $3.50$ & $5.88$\\
\cline{2-6}
$\cal{P}_{\mu^{\mp}}$
   & ${\pm}0.3$& $6.01$ & $3.88$ & $3.05$ & $7.66$\\
\cline{2-6}
   & ${\pm}1$  & $3.92$ & $5.98$ & $1.99$ & $11.8$\\ 
\hline\hline
\end{tabular}
\caption{\it{ 
Neutrino and antineutrino charged currents interaction rates for L = 732, 
3500 and 7332 km per 40 kT and per 5 operational years when $2 \times 10^{20}$ 
muons decay in the straight section of the storage ring. These fluxes have been 
averaged over an angular divergence of 0.1 mr.}}
\label{tab:tab3}
\end{table}

Tab.~\ref{tab:tab3} contains the numerical results for the number of CC interaction rates for 
$e^\pm$ and $\mu^\mp$ fluxes in a $\mu^\mp$ beam in a 40 kT detector with $n_\mu = 2 \times 10^{20}$ 
useful muons per year per 5 operational years. These results represent the number of leptons 
of a given flavour observed at the detector in case no neutrino oscillation 
occurs, neglecting detection efficiencies.

\newpage

{\Large{\bf Appendix B:  Oscillated statistics}} 

As an illustration, we give the oscillated fluxes for three values of the atmospheric mass difference, 
$\Delta m^2_{23} = 2, \, 4, \, 6 \times 10^{-3}$ eV$^2$. The rest of the leptonic parameters are  $\Delta m^2_{12}=  10^{-4}$ eV$^2$, 
$\tetatt= 45^\circ$, $\tetaot= 13^\circ$ and $\theta_{12}=22.5^\circ$. 
The matter parameter is taken to be $A = 1.1 \times 10^{-4}$ eV$^2/$GeV for the baselines of 
$L = $ 732 km and $L = $ 3500 km, while $A = 1.5 \times 10^{-4}$ eV$^2/$GeV for $L =$ 7332 km. 
The in principle measurable quantities are the number of leptons of a given flavour and charge reaching 
the detector at a given baseline. 

Tab.~\ref{tab:lept732mm} shows the total number of leptons of different flavours 
($\bar \nu_e \raw \bar \nu_\mu \raw \mu^+$,$\nu_\mu \raw \nu_e \raw e^-$, 
$\bar \nu_e \raw \bar \nu_\tau \raw \tau^+$ and $\nu_\mu \raw \nu_\tau \raw \tau^-$ for a 
$\mu^-$ beam) at 732 km. Tab.~\ref{tab:lept732mp} shows the analogous results from 
$\mu^+$ decays. Detection efficiencies are not included. 
The whole exercise is repeated for the baselines $L = $ 3500 km and $ L = $ 7332 km
in Tabs.~\ref{tab:lept3500mm}, \ref{tab:lept3500mp} and \ref{tab:lept7332mm}, 
\ref{tab:lept7332mp}, respectively.

\begin{table}[t]
\vspace{2.5cm}
\centering
\begin{tabular}{||c|c||c||c||c||c||c||c||}
\hline\hline
$\cal{P}_{\mu^{-}}$ & $\Delta m_{23}^2$ & $N_{\mu^-}/10^5$ &$N_{e^+}/10^5$&$N_{\mu^+}/10^3$ & $N_{e^-}/10^3$ & $N_{\tau^+}/10^3$ & $N_{\tau^-}/10^4$\\
\hline\hline

     & 0.002 & 691 & 300 & 13.8 & 24.1 & 14.3 & 20.6 \\
\cline{2-8}
$0$     
     & 0.004 & 684 & 299 & 51.3 & 91.7 & 52.9 & 80.9 \\
\cline{2-8}
     & 0.006 & 673 & 298 & 110 & 201 & 113 & 178 \\

\hline\hline  
     & 0.002 & 601 & 390 & 18.0 & 22.0 & 18.6 & 19.4 \\
\cline{2-8}
$0.3$  
     & 0.004 & 595 & 389 & 66.7 & 83.8 & 68.8 & 76.1 \\
\cline{2-8}
     & 0.006 & 584 & 387 & 143 & 184 & 147 & 167  \\

\hline\hline  
     & 0.002 & 391 & 600 & 27.7 & 18.6 & 28.6 & 16.5 \\
\cline{2-8}
$1$  
     & 0.004 & 387 & 598 & 103 & 70.8 & 106 & 64.2  \\
\cline{2-8}
     & 0.006 & 378 & 596 & 220 & 155 & 226  & 140 \\
\hline\hline 
\end{tabular}
\caption{\it  
Calculated charged currents event rates for $\mu^-$ beam assuming neutrino
oscillations in a 40 kT detector, for a L = 732 km baseline as a function of 
$E_\mu$, $\Delta m_{23}^2$, for different polarizations of the parent muon. We 
have considered $1 \times 10^{21}$ negative muons decays ($2 \times 10^{20}$ 
useful muons/year $\times$ 5 operational years).}
\label{tab:lept732mm}
\end{table}

\begin{table}[bht]
\vspace{2.5cm}
\centering
\begin{tabular}{||c|c||c||c||c||c||c||c||}
\hline\hline
$\cal{P}_{\mu^{+}}$ & $\Delta m_{23}^2$ & $N_{\mu^+}/10^5$ &$N_{e^-}/10^5$&$N_{\mu^-}/10^3$ & $N_{e^+}/10^3$ & $N_{\tau^-}/10^3$ & $N_{\tau^+}/10^4$\\
\hline\hline

     & 0.002 & 351 & 590 & 28.9 & 11.7 & 28.0 & 10.5 \\
\cline{2-8}
$0$     
     & 0.004 & 348 & 588 & 110 & 43.6 & 106 & 41.4 \\
\cline{2-8}
     & 0.006 & 342 & 586 & 239 & 94.1 & 232 & 91.5 \\

\hline\hline  
     & 0.002 & 305 & 767 & 37.5 & 11.2 & 36.4 & 9.86 \\
\cline{2-8}
$-0.3$  
     & 0.004 & 302 & 765 & 142 & 41.9 & 138 & 38.7 \\
\cline{2-8}
     & 0.006 & 297 & 762 & 311 & 90.1 & 302 & 85.4  \\

\hline\hline  
     & 0.002 & 200 & 1180 & 57.8 & 9.52 & 56.0 & 8.36 \\
\cline{2-8}
$-1$  
     & 0.004 & 196 & 1170 & 219 & 35.2 & 212 & 32.7  \\
\cline{2-8}
     & 0.006 & 192 & 1172 & 478 & 75.3 & 464  & 71.6 \\
\hline\hline 
\end{tabular}
\caption{\it  
Calculated charged currents event rates for $\mu^+$ beam assuming neutrino 
oscillations in a 40 kT detector, for a L = 732 km baseline as a function of 
$E_\mu$, $\Delta m_{23}^2$, for different polarizations of the parent muon. 
We have considered $1 \times 10^{21}$ positive muons decays ($2 \times 10^{20}$ 
useful muons/year $\times$ 5 operational years).}
\label{tab:lept732mp}
\end{table}

\newpage

\begin{table}[t]
\vspace{2.5cm}
\centering
\begin{tabular}{||c|c||c||c||c||c||c||c||}
\hline\hline
$\cal{P}_{\mu^{-}}$ & $\Delta m_{23}^2$ & $N_{\mu^-}/10^4$ &$N_{e^+}/10^4$&$N_{\mu^+}/10^2$ & $N_{e^-}/10^3$ & $N_{\tau^+}/10^2$ & $N_{\tau^-}/10^3$\\
\hline\hline

     & 0.002 & 282 & 130 & 65.8 & 21.1 & 75.3 & 187 \\
\cline{2-8}
$0$     
     & 0.004 & 232 & 128 & 163 & 81.9 & 187 & 631 \\
\cline{2-8}
     & 0.006 & 170 & 126 & 227 & 174 & 259 & 1150 \\

\hline\hline  
     & 0.002 & 244 & 169 & 85.5 & 17.9 & 97.9 & 176 \\
\cline{2-8}
$0.3$  
     & 0.004 & 198 & 166 & 212 & 70.5 & 243 & 585\\
\cline{2-8}
     & 0.006 & 143 & 164 & 295 & 150 & 337 & 1054 \\

\hline\hline  
     & 0.002 & 156 & 260 & 132 & 15.3 & 151 & 145 \\
\cline{2-8}
$1$  
     & 0.004 & 120 & 256 & 327 & 58.7 & 373 & 463  \\
\cline{2-8}
     & 0.006 & 81.0 & 253 & 454 & 121 & 518  & 792\\
\hline\hline 

\end{tabular}
\caption{\it 
Calculated charged currents event rates for $\mu^-$ beam assuming neutrino 
oscillations in a 40 kT detector, for a L = 3500 km baseline as a function of 
$E_\mu$, $\Delta m_{23}^2$, for different polarizations of the parent muon. 
We have considered $1 \times 10^{21}$ negative muons decays ($2 \times 10^{20}$ 
useful muons/year $\times$ 5 operational years).}
\label{tab:lept3500mm}
\end{table}

\begin{table}[bht]
\vspace{2.5cm}
\centering
\begin{tabular}{||c|c||c||c||c||c||c||c||}
\hline\hline
$\cal{P}_{\mu^{+}}$ & $\Delta m_{23}^2$ & $N_{\mu^+}/10^4$ &$N_{e^-}/10^4$&$N_{\mu^-}/10^3$ & $N_{e^+}/10^2$ & $N_{\tau^-}/10^3$ & $N_{\tau^+}/10^3$\\
\hline\hline

     & 0.002 & 143 & 254 & 25.7 & 60.4 & 22.9 & 100 \\
\cline{2-8}
$0$     
     & 0.004 & 118 & 240 & 96.4 & 164 & 88.1 & 348 \\
\cline{2-8}
     & 0.006 & 86.0 & 221 & 195 & 249 & 181 & 655 \\

\hline\hline  
     & 0.002 & 124 & 330 & 33.4 & 62.4 & 29.8 & 92.7\\
\cline{2-8}
$-0.3$  
     & 0.004 & 100 & 312 & 125 & 165 & 115 & 319 \\
\cline{2-8}
     & 0.006 & 72.4 & 287 & 254 & 246 & 235 & 592  \\

\hline\hline  
     & 0.002 & 79.3 & 507 & 51.4 & 50.2 & 45.8 & 77.0 \\
\cline{2-8}
$-1$  
     & 0.004 & 60.8 & 480 & 193 & 124 & 176 & 254  \\
\cline{2-8}
     & 0.006 & 40.8 & 441 & 390 & 173 & 362  & 450 \\
\hline\hline 
\end{tabular}
\caption{\it 
Calculated charged currents event rates for $\mu^+$ beam assuming neutrino 
oscillations in a 40 kT detector, for a L = 3500 km baseline as a function of 
$E_\mu$, $\Delta m_{23}^2$, for different polarizations of the parent muon. 
We have considered $1 \times 10^{21}$ positive muons decays ($2 \times 10^{20}$ 
useful muons/year $\times$ 5 operational years).}
\label{tab:lept3500mp}
\end{table}

\newpage

\begin{table}[t]
\vskip 2.5cm
\centering
\begin{tabular}{||c|c||c||c||c||c||c||c||}
\hline\hline
$\cal{P}_{\mu^{-}}$ & $\Delta m_{23}^2$ & $N_{\mu^-}/10^3$ &$N_{e^+}/10^3$&$N_{\mu^+}/10^2$ & $N_{e^-}/10^3$ & $N_{\tau^+}/10^2$ & $N_{\tau^-}/10^3$\\
\hline\hline

     & 0.002 & 527 & 299 & 2.38 & 3.97 & 1.94 & 159 \\
\cline{2-8}
$0$     
     & 0.004 & 256 & 296 & 16.7 & 28.2 & 15.1 & 406 \\
\cline{2-8}
     & 0.006 & 89.1 & 292 & 35.8 & 78.8 & 33.5 & 523 \\

\hline\hline  
     & 0.002 & 451 & 388 & 3.10 & 3.95 & 2.52 & 146 \\
\cline{2-8}
$0.3$  
     & 0.004 & 212 & 385 & 21.6 & 27.3 & 19.6 & 362 \\
\cline{2-8}
     & 0.006 & 75.1 & 380 & 46.6 & 74.1 & 43.6 & 452  \\

\hline\hline  
     & 0.002 & 273& 597 & 4.77 & 3.89 & 3.88 & 115 \\
\cline{2-8}
$1$  
     & 0.004 & 110 & 592 & 33.3 & 25.2 & 30.1 & 257  \\
\cline{2-8}
     & 0.006 & 42.5 & 584 & 71.6 & 63.3 & 67.0  & 287 \\
\hline\hline 
\end{tabular}
\caption{\it 
Calculated charged currents event rates for $\mu^-$ beam assuming neutrino 
oscillations in a 40 kT detector, for a L = 7332 km baseline as a function of 
$E_\mu$, $\Delta m_{23}^2$, for different polarizations of the parent muon. 
We have considered $1 \times 10^{21}$ negative muons decays ($2 \times 10^{20}$ 
useful muons/year $\times$ 5 operational years). }
\label{tab:lept7332mm}
\end{table}

\begin{table}[bht]
\vskip 2.5cm
\centering
\begin{tabular}{||c|c||c||c||c||c||c||c||}
\hline\hline
$\cal{P}_{\mu^{+}}$ & $\Delta m_{23}^2$ & $N_{\mu^+}/10^3$ &$N_{e^-}/10^3$&$N_{\mu^-}/10^3$ & $N_{e^+}/10^2$ & $N_{\tau^-}/10^3$ & $N_{\tau^+}/10^3$\\
\hline\hline

     & 0.002 & 265 & 577 & 6.06 & 1.16 & 5.84 & 85.6 \\
\cline{2-8}
$0$     
     & 0.004 & 125 & 513 & 38.1 & 11.5 & 37.8 & 225 \\
\cline{2-8}
     & 0.006 & 52.8 & 399 & 94.8 & 31.2 & 95.0 & 295 \\

\hline\hline  
     & 0.002 & 227 & 750 & 7.88 & 1.20 & 7.59 & 78.6 \\
\cline{2-8}
$-0.3$  
     & 0.004 & 104 & 667 & 49.5 & 11.0 & 49.1 & 201\\
\cline{2-8}
     & 0.006 & 47.6 & 519 & 123 & 28.5 & 123 & 255 \\

\hline\hline  
     & 0.002 & 137 & 1150 & 12.1 & 129 & 11.7 & 62.4 \\
\cline{2-8}
$-1$  
     & 0.004 & 54.8 & 1030 & 76.2 & 10.0 & 75.6 & 144  \\
\cline{2-8}
     & 0.006 & 35.7 & 798 & 190 & 22.3 & 190  & 161 \\
\hline\hline
\end{tabular}
\caption{\it 
Calculated charged currents event rates for $\mu^+$ beam assuming neutrino 
oscillations in a 40 kT detector, for a L = 7332 km baseline as a function of 
$E_\mu$, $\Delta m_{23}^2$, for different polarizations of the parent muon. 
We have considered $1 \times 10^{21}$ positive muons decays ($2 \times 10^{20}$ 
useful muons/year $\times$ 5 operational years). }
\label{tab:lept7332mp}
\end{table}

\newpage

{\Large{\bf Appendix C:  Perturbative Expansion of Oscillation Probabilities}} 

In this Appendix, we describe the perturbative expansion we have performed 
to obtain Eq.~(\ref{approxprob}). The problem is to diagonalize the neutrino 
mass matrix,
\be
M \equiv U \left ( \matrix{ 0 &           0 &      0 \cr 
                            0 & \Delta_{12} &      0 \cr
                            0 &           0 & \delot \cr } \right) U^\dagger +  
           \left ( \matrix{ A & 0 & 0 \cr 
                            0 & 0 & 0 \cr
                            0 & 0 & 0 \cr } \right ) \, ,
\ee
with  $U$ as given in  Eq.~(\ref{CKM}). The exact
diagonalization of this matrix has been done in \cite{zs}.
Since we are interested only in the case in which $\Delta m^2_{12}$ is small, it is 
adequate to use
perturbation theory to compute corrections to first order in this quantity. 
This leads to much simpler analytical formulae. 

In the limit $\Delta m^2_{12} = 0$, the diagonalization of this matrix is very simple \cite{Yasuda}, 
\be
\label{diag0}
M_\mp^{(0)} \equiv \bar U_\mp \left ( \matrix{ \frac{\Delta_{13} \pm A - B_\mp}{2} & 0 & 0 \cr 
                                                                                 0 & 0 & 0 \cr
                                               0 & 0 & \frac{\Delta_{13} \pm A + B_\mp}{2} \cr}
\right) \bar U_\mp^\dagger \, . 
\ee 
The matrix of eigenvectors is
\be
\bar U_\mp \equiv U_{23} (\tetatt) U_{13} (\theta_{M_\mp}) \, , 
\ee
where $\theta_{M_\mp}$ is defined by: 
\be
\tan 2 \theta_{M_\mp} \equiv \frac{\delot \sin 2 \tetaot}{\delot \cos 2 \tetaot \mp A} \, . 
\ee
$\theta_{M_\mp}$ is to be taken in the first (second) quadrant
if $\delot \,\cos 2 \tetaot \mp A$ is positive (negative).

At first order in $\Delta_{12}$, the perturbation to Eq.~(\ref{diag0}) 
(in the basis of non-perturbated eigenvectors) is:
\be
\label{diag1}
M_\mp^{(1)} \equiv \bar U_\mp^\dagger U \left ( \matrix{ 0 &           0 & 0 \cr 
                                                         0 & \Delta_{12} & 0 \cr
                                                         0 &           0 & 0 \cr } \right ) 
         U^\dagger \bar U_\mp. 
\ee
The eigenvalues at first order in $\Delta_{12}$ are:
\bea
\lambda^{(1)}_1&=& \lambda^{(0)}_1 + s^2_{12} \Delta_{12} \cos^2 \bar \theta_{M_\mp} \, , \nn \\
\lambda^{(1)}_2&=& \lambda^{(0)}_2 + c^2_{12} \Delta_{12} \, , \nn \\
\lambda^{(1)}_3&=& \lambda^{(0)}_3 + s^2_{12} \Delta_{12} \sin^2 \bar \theta_{M_\mp} \, , 
\eea
The corresponding (not normalized) eigenvectors are, 
\bea
v^{(1)}_1 & = & \left(1 ,
               \ \frac{ \sin 2 \theta_{12} \cos \bar \theta_{M_\mp} \Delta_{12} }{
                        \delot + A - B_\mp} ,
               \ \frac{s^2_{12} \Delta_{12} \sin \bar \theta_{M_\mp} 
                        e^{\mp i\delta}}{2B_\mp} \right ) \, , \nn\\
v^{(1)}_2 & = & \left ( \frac{ \sin 2 \theta_{12} \cos \bar \theta_{M_\mp} \Delta_{12} }{
                       -\delot - A + B_\mp} ,
               \  1,
               \ \frac{ \sin 2 \theta_{12} \sin \bar \theta_{M_\mp} 
                        e^{\mp i\delta} \Delta_{12}}{\delot + A + B_\mp} \right ) \, , \nn \\
v^{(1)}_3 & = & \left ( \frac{-s^2_{12} \sin 2 \bar \theta_{M_\mp} \Delta_{12} 
                        e^{\pm i\delta}}{2B_\mp} ,
               \ \frac{-\sin 2 \theta_{12} \sin \bar \theta_{M_\mp} \Delta_{12} e^{\pm i\delta}}{
                        \Delta_{31} + A + B_\mp} ,\ 1 \right ), \,
\eea
where $\bar \theta_{M_\mp} \equiv \tetaot - \theta_{M_\mp}$. 

With these results, it is easy to compute the oscillation probabilities, 
keeping consistently terms up to first order in $\Delta_{12}$. We obtain: 
\bea
& & P_{\nu_e \nu_\mu ( \bar \nu_e \bar \nu_\mu) } = 
       s^2_{23} \sin^2 (2 \theta_{M_\mp}) \sin^2 \left( \frac{ B_\mp \, L}{2} \right ) -
       \nn\\
& &    s^2_{23} s^2_{12} \left [ \sin ( 4 \theta_{M_\mp} ) \sin ( 2 \bar \theta_{M_\mp} )
       \sin^2 \left ( \frac{B_\mp \, L}{2} \right ) 
                      \frac{\Delta_{12}}{ B_\mp} + \sin^2 ( 2 \theta_{M_\mp} )
       \cos ( 2 \bar \theta_{M_\mp} ) \sin ( B_\mp \, L ) 
                      \frac{ \Delta_{12}\, L}{2} \right ]  + \nn \\
& &    \sin ( 2 \theta_{12} ) \sin ( 2 \tetatt ) \sin ( 2 \theta_{M_\mp} ) 
       \sin \left ( \frac{B_\mp \, L}{2} \right ) \Delta_{12} \times \nn \\
& &    \left [ \sin \left ( \frac{ \lambda^{(0)}_1 \, L}{2} \right )
               \cos \left ( \pm \delta - \frac{ \lambda^{(0)}_3 \, L}{2} \right ) 
                    \left ( \frac{ \cos \theta_{M_\mp} \cos \bar \theta_{M_\mp}}{\lambda^{(0)}_1} - 
                            \frac{ \sin \theta_{M_\mp} \sin \bar \theta_{M_\mp}}{\lambda^{(0)}_3} 
             \right ) - \right . \nn \\ 
& & \label{olgaprob1} \left .
        \sin \theta_{M_\mp} \sin \bar \theta_{M_\mp} \cos \delta 
        \sin \left ( \frac{B_\mp \, L}{2} \right ) \frac{1}{\lambda^{(0)}_3} \right ] \, ,
\eea 
\bea
& &P_{\nu_e \nu_\tau (\bar \nu_e \bar \nu_\tau) } = 
       c^2_{23} \sin^2 (2 \theta_{M_\mp} ) \sin^2 \left ( \frac{B_\mp \, L}{2} \right ) -
\nn \\
& &    c^2_{23} s^2_{12} \left [ \sin ( 4 \theta_{M_\mp} ) \sin ( 2 \bar \theta_{M_\mp} )
       \sin^2 \left ( \frac{B_\mp \, L}{2} \right ) \frac{\Delta_{12}}{B_\mp} +
       \sin^2 ( 2 \theta_{M_\mp} )
       \cos(2 \bar \theta_{M_\mp}) \sin(B_\mp \, L) \frac{\Delta_{12} \, L}{2} \right] - 
       \nn \\
& &   \sin(2\theta_{12})\ \sin(2\tetatt)\ \sin(2\theta_{M_\mp})\ 
      \sin \left ( \frac{B_\mp \, L}{2} \right ) \Delta_{12} \times \nn \\
& &   \left [ \sin \left ( \frac{ \lambda^{(0)}_1 \, L}{2} \right ) 
      \cos \left ( \pm \delta - \frac{ \lambda^{(0)}_3 \, L}{2} \right ) 
           \left ( \frac{ \cos \theta_{M_\mp} \cos \bar \theta_{M_\mp}}{ \lambda^{(0)}_1} -
                   \frac{ \sin \theta_{M_\mp} \sin \bar \theta_{M_\mp}}{ \lambda^{(0)}_3} 
            \right ) - \right . \nn \\ 
& & \label{olgaprob2} \left . 
        \sin \theta_{M_\mp} \sin \bar \theta_{M_\mp} \cos \delta
        \sin \left ( \frac{B_\mp \, L}{2} \right ) \frac{1}{ \lambda^{(0)}_3} \right ] \, ,
\eea 
\bea
& &P_{\nu_e \nu_e (\bar \nu_e \bar \nu_e ) } = 
       1 - \sin^2 (2 \theta_{M_\mp} ) \sin^2 \left( \frac{B_\mp \, L}{2} \right ) + \nn \\
& &  s^2_{12} \left [ \sin ( 4 \theta_{M_\mp}) \sin ( 2 \bar \theta_{M_\mp}) 
             \sin^2 \left ( \frac{B_\mp \, L}{2} \right ) \frac{ \Delta_{12}}{B} + 
             \sin^2 ( 2 \theta_{M_\mp} ) \cos ( 2 \bar \theta_{M_\mp} ) 
             \sin ( B_\mp \, L)\frac{\Delta_{12} \, L}{2} \right ] \, . 
\nn \\
\nn \\
\label{olgaprob3}
\eea
These formulae are valid for all values of $\tetaot$ and to first 
order in $\Delta_{12}$. It is rather straightforward to check that they reduce 
to the vacuum result for $A\raw 0$. 

In section 3, we have further considered an expansion in which not only 
$\Delta_{12}$ but also $\tetaot$ are small. We have kept terms
up to second order: i.e. ${\cal O}(\Delta^2_{12} \tetaot^0)$, 
${\cal O}(\Delta_{12} \tetaot)$ and ${\cal O}( \Delta_{12}^0 \tetaot^2 )$.
The latter two can be obtained
from Eqs.~(\ref{olgaprob1},\ref{olgaprob2}) and (\ref{olgaprob3}), 
by performing an expansion in $\tetaot$. 
On the other hand, the terms of ${\cal O}(\Delta_{12}^2 \tetaot^0)$ are absent 
in that approximation, they appear at next order in the expansion.
They can be easily obtained, though, starting  directly from the diagonalization
 of the mass matrix with $\tetaot = 0$:  
\be
M \equiv U_{23} (\tetatt) U_{12} (\theta'_{M_\mp}) 
 \left(\matrix{ \frac{ \pm A + \Delta_{12} - C_\mp}{2} & 0 &   0 \cr 
  0 & \frac{\pm A +\Delta_{12} + C_\mp}{2} &      0 \cr
  0 & 0 & \delot \cr } \right) 
  U_{12}(\theta'_{M_\mp})^\dagger U_{23} (\tetatt)^\dagger,
\ee
with $C_\mp \equiv \sqrt{\Delta_{12}^2 + A^2 \mp 2 A \Delta_{12} \cos 2 \theta_{12}}$ and
\be
\sin 2 \theta'_{M_\mp} \equiv - \frac{\Delta_{12} \sin 2 \theta_{12}}{C_\mp}.
\ee
The expansion of the corresponding probabilities to second order in $\Delta_{12}$ gives, 
\bea
P_{\nu_e \nu_\mu  (\bar \nu_e \bar \nu_\mu)} & = & 
   c^2_{23} \sin^2 2 \theta_{12} 
   \left ( \frac{\Delta_{12}}{A} \right )^2 \sin^2 \left ( \frac{A \, L}{2} \right ) \nn \\
P_{\nu_e \nu_\tau (\bar \nu_e \bar \nu_\tau)} & = & 
   s^2_{23} \sin^2 2 \theta_{12} 
   \left ( \frac{\Delta_{12}}{A} \right )^2 \sin^2 \left ( \frac{A \, L}{2} \right ) \nn \\
P_{\nu_e \nu_e (\bar \nu_e \bar \nu_e)} & = & 
1 -  \sin^2 2 \theta_{12} 
   \left ( \frac{\Delta_{12}}{A} \right )^2 \sin^2 \left ( \frac{A \, L}{2} \right ) \, .  
\eea
From the first of these equations we obtain the term in $\Delta_{12}^2$ of 
Eq.~(\ref{approxprob}).

\end{document}